\documentclass[a4paper,11pt]{article}
\usepackage{epsfig,amssymb,amsfonts,amsmath,mathtools,bm,color}

\usepackage{jheppub}

\newcommand{\be}{\begin{equation}}
\newcommand{\ee}{\end{equation}}
\newcommand{\bea}{\begin{eqnarray}}
\newcommand{\eea}{\end{eqnarray}}
\newcommand{\beas}{\begin{eqnarray*}}
\newcommand{\eeas}{\end{eqnarray*}}
\newcommand{\vep}{{\bm p}}

\newcommand{\veq}{{\bm q}}

\newcommand{\veA}{{\bm A}}
\newcommand{\veep}{{\bm \epsilon}}

\newcommand{\ds}{\displaystyle}
\newcommand{\vetau}{{\bm \tau}}

\graphicspath{{Figs.dir/}}
\title{\boldmath Spin partners of the $Z_b(10610)$ and $Z_b(10650)$ revisited}

\author[a,b]{V. Baru}
\author[a]{E. Epelbaum}
\author[a]{A. A. Filin}
\author[c]{C. Hanhart}
\author[b,d,e]{A. V. Nefediev}

\affiliation[a]{Institut f\"ur Theoretische Physik II, Ruhr-Universit\"at Bochum, D-44780 Bochum, Germany}
\affiliation[b]{Institute for Theoretical and Experimental Physics, B. Cheremushkinskaya 25, 117218 Moscow, Russia}
\affiliation[c]{Forschungszentrum J\"ulich, Institute for Advanced Simulation, Institut f\"ur Kernphysik and
J\"ulich Center for Hadron Physics, D-52425 J\"ulich, Germany}
\affiliation[d]{National Research Nuclear University MEPhI, 115409, Kashirskoe highway 31, Moscow, Russia}
\affiliation[e]{Moscow Institute of Physics and Technology, 141700, Institutsky lane 9, Dolgoprudny, Moscow Region, Russia}

\emailAdd{vadimb@tp2.rub.de}
\emailAdd{evgeny.epelbaum@ruhr-uni-bochum.de}
\emailAdd{arseniy.filin@ruhr-uni-bochum.de}
\emailAdd{c.hanhart@fz-juelich.de}
\emailAdd{nefediev@itep.ru}

\abstract{
We study the implications of the heavy-quark spin symmetry for the possible spin partners of the exotic
states $Z_b(10610)$ and $Z_b(10650)$ in the spectrum
of bottomonium. We formulate and solve numerically the coupled-channel equations for the $Z_b$ states
that allow for a dynamical generation of these states as hadronic molecules.
The force includes short-range contact terms and the one-pion exchange potential, both treated fully nonperturbatively.
The strength of the potential at leading order is fixed completely by the pole positions of the $Z_b$ states
so that the mass and the most prominent contributions to the width of the isovector heavy-quark spin partner states
$W_{bJ}$ with the quantum numbers $J^{++}$ ($J=0,1,2$) come out as predictions.
{In particular, we predict the existence of an
isovector $2^{++}$ tensor state lying a few MeV below the $B^*\bar{B}^*$ threshold which should be detectable in the experiment.}
Since the accuracy of the present experimental data does not allow
one to fix the pole positions of the $Z_b$'s reliably enough,
we also study the pole trajectories of their spin partner states as functions
of the $Z_b$ binding energies. It is shown that, once the heavy-quark spin symmetry
  is broken by the physical $B$ and $B^*$ mass difference, especially the pion
tensor force has a significant impact on the location of the partner states clearly
demonstrating the need of a coupled-channel treatment of pion dynamics to understand the spin multiplet pattern of hadronic molecules.
}


\begin{document}
\maketitle
\flushbottom

\section{Introduction}

The experimental discovery of the charmonium-like state $X(3872)$ by the Belle Collaboration in
2003~\cite{Choi:2003ue} inaugurated a new era in  hadron spectroscopy. A lot of new
candidates for exotic states have been discovered since then
in the spectrum of both charmonium and bottomonium --- for a review see, for example,
Refs.~\cite{Brambilla:2010cs,Eidelman:2016qad}. Among those one should especially mention the isovector
$Z_b(10610)$ and $Z_b(10650)$ resonances (for brevity, hereinafter often referred to as $Z_b$ and $Z_b'$, respectively)
\cite{Belle:2011aa,Adachi:2012cx,Garmash:2015rfd}.
Their  signals are seen in 7 channels, namely
\begin{eqnarray}
\Upsilon(10860)&\to&\pi Z_b^{(\prime)}\to \pi B^{(*)}\bar{B}^*,\nonumber \\
\Upsilon(10860)&\to&\pi Z_b^{(\prime)}\to\pi\pi\Upsilon(nS),\quad n=1,2,3,\label{chains}\\
\Upsilon(10860)&\to&\pi Z_b^{(\prime)}\to\pi\pi h_b(mP),\quad m=1,2\nonumber \ .
\end{eqnarray}
Since $Z_b$ and $Z_b'$ are charged but decay into  final states containing a $b$ and an anti-$b$ quark 
they must contain at least four quarks and are thus explicitly exotic.
It turned out that the decays of the $Z_b$'s into the open-flavour channels almost exhaust their widths, despite the limited phase spaces.
This feature is to be regarded as a strong evidence for a molecular
nature of the $Z_b$ states~\cite{Bondar:2011ev} --- for a recent review of the theory of hadronic molecules we
refer to Ref.~\cite{Guo:2017jvc}.

The corresponding masses and widths extracted from the Breit-Wigner fits for the data and averaged over the above production and decay channels
are~\cite{Olive:2016xmw}
\begin{eqnarray}
M_{Z_b}=10607.2\pm2.0~\mbox{MeV},\quad \Gamma_{Z_b}=18.4\pm 2.4~\mbox{MeV},\label{Zb1exp}\\
M_{Z_b'}=10652.2\pm 1.5~\mbox{MeV},\quad\Gamma_{Z_b'}=11.5\pm 2.2~\mbox{MeV},\label{Zb2exp}
\end{eqnarray}
to be compared with 10604 MeV and 10649 MeV for the $B\bar B^*$ and the $B^*\bar B^*$ thresholds, respectively.
Therefore, both poles are located within a couple of MeV from the respective threshold.
The $J^{PC}$ quantum numbers of both states were determined as $1^{+-}$~\cite{Collaboration:2011gja} in line
with the expectation of the molecular model.

Within the molecular picture, the wave functions of the $Z_b$ states can be written as~\cite{Bondar:2011ev}
\begin{eqnarray}
\label{Zbmolecule}
&1^+(1^{+-}):~&|Z_b\rangle=-\frac{1}{\sqrt 2}\Bigl[(1^-_{b \bar b}\otimes 0^-_{q \bar q})_{S=1}+
(0^-_{b \bar b}\otimes 1^-_{q \bar q})_{S=1}\Bigr]\label{Zb1},\\ \label{Zb'molecule}
&1^+(1^{+-}):~&|Z_b'\rangle=\frac{1}{\sqrt{2}}\Bigl[(1^-_{b \bar b}\otimes 0^-_{q \bar
q})_{S=1}-(0^-_{b \bar b}\otimes 1^-_{q \bar q})_{S=1}\Bigr], \label{Zb2}
\end{eqnarray}
where we quote the quantum numbers in the form $I^G(J^{PC})$ and, for example, $1^-_{b \bar b}$ denotes the wave function of the $b\bar{b}$ pair with
the total spin 1, and so forth.
Such an identification of the $Z_b$'s allows one to explain the
dipion transition rates from the $\Upsilon(10860)$ to the vector bottomonia $\Upsilon(nS)$ ($n=1,2,3$), which appear to be up to two orders of magnitude
larger than similar transitions among the lower bottomonia $\Upsilon(nS)$ ($n=1,2,3$)~\cite{Garmash:2014dhx}.
In addition, the molecular interpretation of the $Z_b$'s (see Eqs.~(\ref{Zbmolecule})-(\ref{Zb'molecule})) naturally explains that  transitions  such as
$Z_b^{(\prime)}\to\pi h_b(mP)$, which are expected to be suppressed by $(\Lambda_{\rm QCD}/m_b)$ since they involve a change
in the heavy-quark spin, happen at a comparable rate
to  the heavy-quark spin preserving transitions $Z_b^{(\prime)}\to\pi \Upsilon(nS)$~\cite{Belle:2011aa}. 

Molecular states should in general be located below the most relevant threshold and not
above\footnote{Meson-meson dynamics can also lead to above-threshold poles if the interaction is energy-dependent,
but such states are not expected
to be very narrow~\cite{Hanhart:2014ssa}. On the other hand, a tetraquark nature of the states is claimed to result in the poles lying slightly above
threshold~\cite{Esposito:2016itg}.}, which seems to be in conflict with the numbers quoted in Eqs.~(\ref{Zb1exp}) and (\ref{Zb2exp}). However, these numbers
need to be interpreted with care. While they show unambiguously that the poles related to the
$Z_b$ states reside very close to the corresponding open-flavour thresholds, their particular values contain an
uncontrollable intrinsic systematic uncertainty since they were determined from sums of the Breit-Wigner functions that ignore the
nearby thresholds and which, therefore, appear to be in conflict with analyticity and unitarity. 
Also, it is argued, for example, in Ref.~\cite{Chen:2015jgl} (see also Ref.~\cite{Baru:2004xg} for a related discussion in the context 
of $f_0(980)/a_0(980)$) that the branching fractions
extracted from the Breit-Wigner parameterisations do not represent the decay probabilities for near-threshold states.
Thus, a more refined data analysis is required. For example, it is demonstrated in
Ref.~\cite{Cleven:2011gp} that both $Z_b$'s are compatible with bound state poles in the data for the $h_b(mP)\pi$ channels as soon as the energy dependence of their
self-energies is included properly. On the other hand, a combined analysis of the experimental data in all seven channels
listed in Eqs.~(\ref{chains}) consistent with analyticity and unitarity
favours both $Z_b$'s as virtual states located within approximately 1~MeV below the respective
thresholds~\cite{Hanhart:2015cua,Guo:2016bjq}\footnote{While the formalism of these references is
indeed unitary with respect to all seven channels, only the open-flavour and the $h_b(mP)\pi$ channels
were included in the fit, since the $\Upsilon(nS)\pi$ channels call for an additional inclusion of nonresonant production.}. In other words,
the quality of the existing data does not allow one to draw  definite conclusions about the pole locations of these states.
Thus, in the present paper, we investigate the fate of the symmetry partners of the $Z_b$ states
as the binding energies of the latter are varied up to zero. As soon as the \mbox{analysis} of Refs.~\cite{Hanhart:2015cua,Guo:2016bjq} is refined
to include one-pion exchange interactions, the predictions for the partner states would have become possible directly from the fit 
for the experimental line shapes.
However, a detailed investigation of the line shapes, which also calls for an inclusion of the inelastic channels, is delegated to  future studies.

In the limit of an infinite $b$-quark mass, a molecular nature of the $Z_b$ states allows
one to predict the existence of spin partners with the quantum numbers $J^{PC}=J^{++}$, conventionally denoted in the literature as $W_{bJ}$,
with $J=0,1,2$~\cite{Bondar:2011ev,Voloshin:2011qa,Mehen:2011yh} (see also Ref.~\cite{Bondar:2016hva} for a
recent review and for the discussion of the future experimental tasks). In particular, in the notations
of Eqs.~(\ref{Zb1}) and (\ref{Zb2}), the wave functions of the spin partners take the form
\begin{eqnarray}
&1^-(0^{++}):~&|W_{b0}\rangle=\frac12\Bigl[\sqrt{3} (1_{b\bar{b}}\otimes 1_{q\bar{q}})_{S=0}
-(0_{b\bar{b}}\otimes 0_{q\bar{q}})_{S=0}\Bigr],\label{Wb0}\\
&1^-(0^{++}):~&|W_{b0}'\rangle=\frac12\Bigl[(1_{b\bar{b}}\otimes 1_{q\bar{q}})_{S=0}+
\sqrt{3}(0_{b\bar{b}}\otimes 0_{q\bar{q}})_{S=0}\Bigr],\label{Wb0pr}\\[1mm]
&1^-(1^{++}):~&|W_{b1}\rangle=(1_{b\bar{b}}\otimes 1_{q\bar{q}})_{S=1},\label{Wb1}\\[2mm]
&1^-(2^{++}):~&|W_{b2}\rangle=(1_{b\bar{b}}\otimes 1_{q\bar{q}})_{S=2}.\label{Wb2}
\end{eqnarray}

Although a detailed microscopic theory is needed to predict individual decay widths of the $Z_b$ and $W_{bJ}$ states, the spin symmetry allows one to
arrive at various relations between their total widths, for example \cite{Voloshin:2011qa},
\begin{eqnarray}
&\Gamma[Z_b]=\Gamma[Z_b'],&\label{ZZwidth}\\
&\ds\Gamma[W_{b1}]=\Gamma[W_{b2}]=\frac32\Gamma[W_{b0}]- \frac12 \Gamma[W_{b0}'],&
\label{WWwidth}
\end{eqnarray}
and~\cite{Mehen:2011yh}
\begin{equation}
\Gamma[Z_b]=\Gamma[Z_b']=\frac{1}{2}\left(\Gamma[W_{b0}] +\Gamma[W_{b0}']\right).\label{ZWwidth}
\end{equation}

Additional predictions for the partial decay widths into various hidden-bottom final states can be also found in
Refs.~\cite{Voloshin:2011qa,Mehen:2011yh}.

To arrive at the above predictions the mass splitting
\begin{equation}
\delta = m_* - m = 45 \ \mbox{MeV}
\label{deldef}
\end{equation}
was treated as a large scale (compared to the binding energies) which was integrated out. Hereinafter $m$ and $m_*$ denote the $B$ and $B^*$ masses,
respectively. While this assumption is acceptable for the $Z_b$ states
themselves\footnote{The first correction to the leading-order result (\ref{ZWwidth}) is controlled by the parameter $\sqrt{E_B/\delta}$~\cite{Baru:2016iwj}, 
which, for example,
for the binding energy $E_B=5$~MeV yields the uncertainty around 30\%.}, for their spin partner it turns out to be appropriate only
within a truncated scheme when only the $S$-wave interactions are retained.
However, it is the central finding of this paper that, as soon as the one-pion exchange (OPE) is included, the effect
of the mass difference $\delta$ is enhanced significantly via the
strong $S$-$D$ transitions that come with it. This results in  binding energies of the spin partner states
as large as 20 MeV, which are clearly of the order of the spin symmetry violating parameter $\delta$ --- see Eq.~(\ref{deldef}).
While pion exchanges were already included in studies of the
$Z_b$ states before~\cite{Sun:2011uh}, to the best of our knowledge its impact on the spin partners has not been investigated so far.
Furthermore, since, in Ref.~\cite{Dias:2014pva}, the effect of the OPE was claimed to be somewhat diminished if the
one-$\eta$ exchange (OEE) is added, we also investigate the effect of this addition.

It is well known that a nonperturbative inclusion of the OPE interaction in the chiral nuclear effective-field theory
allows one to significantly extend the region of applicability
of the theory --- see, for example, Refs~\cite{Epelbaum:2008ga,Fleming:1999ee}. In charmonium-like systems such as the $X(3872)$,
pionic effects turn out to be important to predict the dynamical properties such as the light-quark mass dependence of
the $X$-pole~\cite{Baru:2013rta,Baru:2015tfa} (for a discussion of the pion mass dependence with
perturbative pions we refer to Ref.~\cite{Jansen:2013cba}) and the decay width
$X(3872) \to D\bar D\pi$~\cite{Baru:2011rs}. Furthermore, it was shown in Ref.~\cite{Baru:2016iwj} that the inclusion of
the OPE interactions had a strong impact on the location of the spin partner states of the $X(3872)$.
In particular, the iterations of the OPE interaction to all orders were found to generate strong coupled-channel effects due to the
then allowed transitions $D^*\bar{D}^*\leftrightarrow D^{(*)}\bar{D}$. As a consequence,
the binding energy and the width of the spin-2 partner of the $X(3872)$ both appeared to be of the order of several dozens of MeV.
Both values were found to be significantly larger than those reported in the literature earlier~\cite{Albaladejo:2015dsa}.

In this paper, a calculation for the spin partners of the $Z_b$ states is presented in which
the pole positions of the $Z_b$'s are treated as input.
The underlying coupled-channel model includes short-range contact interactions, OPE, and OEE
fully iterated to all orders. As our starting point, we assume both $Z_b$'s to be shallow bound states and investigate the
role played by the spin symmetry violation introduced into the system via a nonzero value
of the parameter $\delta$ defined in Eq.~(\ref{deldef}). In addition, we study the behaviour of the spin partners as
the binding energies of the $Z_b$ states are reduced (up to zero values) which allows us to make statements about
the masses of the spin partners when the $Z_b$'s turn to virtual states.
In particular, we find that the tensor spin partner $W_{b2}$ with the quantum numbers $2^{++}$
survives in this limit \emph{as a bound state} and it is expected to produce a visible resonant
structure in the $B\bar B$ and $B\bar B^*$ line shapes a few MeV below the $B^*\bar{B}^*$
threshold.

The key aim of this study is to highlight explicitly the nontrivial impact of the nonperturbative one-pion exchange
on the location of the spin partners of the $Z_b$ states.
For that it appears sufficient to base the study solely on the open-flavour channels introduced below. In particular, we neither
include the inelastic channels ($\pi \Upsilon(nS)$ and $\pi h_b(mP)$) nor do we aim at a high accuracy description of the data. Clearly,
as soon as the analysis of Refs.~\cite{Hanhart:2015cua,Guo:2016bjq} is refined 
to include one-pion exchange interactions, the predictions for the partner states could be refined further.
Such a study, however, is beyond the scope of this paper. 

{The paper is organised as follows. In Sec.~\ref{sec:contact}, we outline various implications of the heavy-quark spin symmetry (HQSS) in the purely contact theory.
Then, in Sec.~\ref{sec:OPE}, we discuss the inclusion of the OPE interaction on top of the contact potential introduced in Sec.~\ref{sec:contact}.
The results obtained in the  theory which incorporates both the contact and the OPE interactions are collected and discussed in Sec.~\ref{sec:results}. A peculiar 
alternative solution found in the contact theory which, however, 
disappears from the theory once the OPE interaction is taken into account is highlighted in Sec.~\ref{sec:comment}. We give an overview 
of the results obtained in the concluding Sec.~\ref{sec:concl}.} 

\section{Contact theory}
\label{sec:contact}

We start from the purely contact theory. The basis of states --- see Ref.~\cite{Nieves:2012tt} ---
can be adapted directly from the
$c$-sector to the $b$-sector,
\begin{eqnarray}
&&0^{++}:\quad\left\{B\bar{B}({^1S_0}),B^*\bar{B}^*({^1S_0})\right\},\nonumber\\
&&1^{+-}:\quad\left\{B\bar{B}^*({^3S_1},-),B^*\bar{B}^*({^3S_1})\right\},\nonumber\\[-3mm]
\label{basisct}\\[-3mm]
&&1^{++}:\quad\left\{B\bar{B}^*({^3S_1},+)\right\},\nonumber\\
&&2^{++}:\quad\left\{B^*\bar{B}^*({^5S_2})\right\}.\nonumber
\end{eqnarray}
Here, the individual partial waves are labelled as $^{2S+1}L_J$ with $S$, $L$, and $J$ denoting
the total spin, the angular momentum, and the total momentum of the two-meson system, respectively. The C-parity eigenstates are defined
as\footnote{In what follows, $B\bar{B}^*$ is used as a shorthand notation for the combination with the appropriate C-parity; the latter
may be stated explicitly if necessary, like in Eq.~(\ref{basisct}).}
\begin{equation}
B\bar{B}^*(\pm)=\frac{1}{\sqrt{2}}\left(B\bar{B}^*\pm B^*\bar{B}\right)
\end{equation}
and comply with the convention for the C-parity transformation $\hat{C}{\cal M}=\bar{\cal M}$.

In the basis of Eq.~(\ref{basisct}) and for a given set of the $J^{PC}$ quantum numbers
the leading-order EFT potentials $V^{(JPC)}_{\rm LO}$
which respect the heavy-quark spin symmetry read~\cite{AlFiky:2005jd,Nieves:2012tt,Valderrama:2012jv}
\begin{eqnarray}\label{C0++}
&&V_{\rm LO}^{(0{++})}=
\begin{pmatrix}
C_{1a} & -\sqrt{3}C_{1b} \\
-\sqrt{3}C_{1b} & C_{1a}-2C_{1b}
\end{pmatrix}
\equiv\frac14
\begin{pmatrix}
3C_1+C_1' & -\sqrt{3}(C_1-C_1') \\
-\sqrt{3}(C_1-C_1') & C_1+3C_1'
\end{pmatrix},
\label{Vct0++}\\
&&V_{\rm LO}^{(1{+-})}=
\begin{pmatrix}
C_{1a}-C_{1b} & 2C_{1b} \\
2C_{1b} & C_{1a}-C_{1b}
\end{pmatrix}
\equiv\frac12
\begin{pmatrix}
C_1+C_1' & C_1-C_1' \\
C_1-C_1' & C_1+C_1'
\end{pmatrix},
\label{Vct1+-}\\
&&V_{\rm LO}^{(1{++})}=C_{1a}+C_{1b}\equiv C_1 \label{eq:contact2-a},\label{Vct1++} \\
&&V_{\rm LO}^{(2{++})}=C_{1a}+C_{1b}\equiv C_1 \label{eq:contact2-b},\label{Vct2++}
\end{eqnarray}
where $\{C_{1a},C_{1b}\}$ and $\{C_1,C_1'\}$ are two alternative sets of low-energy constants (contact terms).

With the help of the appropriate unitary transformations, the potential matrices (\ref{Vct0++}) and (\ref{Vct1+-})
can be diagonalised to take the form $\mbox{diag}(C_1,C_1')$.
Therefore, in the strict heavy-quark limit ($\delta=0$) and in the
purely contact theory, if the potential $C_1^{(\prime)}$ is strong enough to bind the system in one channel, it is inevitably strong enough to
produce bound state(s) in the sibling channel(s). Furthermore, since the binding energies in different channels are governed by the same combinations
of the contact terms, the corresponding molecular states appear to be degenerate in mass. This observation allowed the authors
of Refs.~\cite{Hidalgo-Duque:2013pva,Baru:2016iwj} to predict, in
the strict HQSS limit, the existence of multiple degenerate isosinglet states in the $c$-sector. Obviously, this conclusion trivially translates
into the $b$-sector and holds for the isovector states, too. In particular,  one  finds
\begin{equation}
E_B^{(0)}[W_{b0}]=E_B^{(0)}[W_{b1}]=E_B^{(0)}[W_{b2}]=E_B^{(0)}[Z_b]\quad\mbox{and}\quad E_B^{(0)}[W_{b0}^\prime]=E_B^{(0)}[Z_b'],
\label{HQSSlimit}
\end{equation}
where the superscript ${(0)}$ indicates the exact HQSS limit which implies the relation
\begin{equation}
\delta\ll E_B\ll m
\label{rel1}
\end{equation}
between the spin symmetry violating parameter $\delta$, the typical binding energy $E_B$ in the $Z_b$ and $W_{bJ}$ states, and the heavy-meson mass.
Therefore, the input provided by the masses
of the two $Z_b$ states is sufficient to fix the parameters of the contact interaction at leading order.
In Refs.~\cite{HidalgoDuque:2012pq,Bondar:2011ev,Voloshin:2011qa,Mehen:2011yh} a different relation between the relevant scales of the problem was assumed,
\begin{equation}
E_B\ll\delta\ll m,
\label{rel2}
\end{equation}
while the contact interaction at this order was still treated as spin symmetric --- see Eqs.~(\ref{C0++})-(\ref{Vct2++}). Then,
the degeneracy (\ref{HQSSlimit}) is lifted and instead one arrives at some relations between the binding momenta of the partners,
\begin{eqnarray}
\gamma_{\scriptscriptstyle Z_b}=\gamma_{\scriptscriptstyle Z_b'}, \quad 
\gamma_{\scriptscriptstyle W_{b1}}=\gamma_{\scriptscriptstyle W_{b2}}, \quad 
\gamma_{\scriptscriptstyle W_{b0}}=\frac{\gamma_{\scriptscriptstyle Z_b}+\gamma_{\scriptscriptstyle W_{b1}}}{2},
\quad \gamma_{\scriptscriptstyle W_{b0}'}=\frac{3\gamma_{\scriptscriptstyle Z_b}-\gamma_{\scriptscriptstyle W_{b1}}}{2},
\label{HQSSdeltaLO}
\end{eqnarray}
which, in particular, yield the relations (\ref{WWwidth})-(\ref{ZWwidth}) for the widths. It should be noted that, to this order,
the states $Z_b'$ and $W_{b0}'$ cannot yet decay into the (open) channels $B\bar B^*$ and $B\bar B$, respectively, and the widths
(\ref{WWwidth})-(\ref{ZWwidth}) originate entirely from inelastic transitions.
As was shown in Ref.~\cite{Baru:2016iwj}, already at order $O(\delta)$, relations (\ref{HQSSdeltaLO}) acquire corrections linear with the cutoff introduced to regularise
the divergent integrals in the Lippmann-Schwinger equations. Then, to absorb this cutoff dependence, the renormalisation in principle calls for the presence of
HQSS violating contact interactions, in addition to the HQSS preserving potentials~(\ref{C0++})-(\ref{Vct2++}).
Since the HQSS violating counter terms are not known, as an alternative approach,
one may vary the cutoff in some reasonable range to estimate the uncertainty in the position of the partner states due to the omission of
these interactions.
In practice, the cutoff dependence appears to be very marginal (at least in the pionless theory) which can be, in part,
justified by the large $b$-quark mass (much larger than the characteristic
cutoff varied in the range from the pion mass and up to the chiral symmetry breaking scale $\sim$ 1 GeV) and by
the additional suppression of such terms by a factor $\gamma/m$ (at least for the uncoupled partial waves)~\cite{Baru:2016iwj}.

Note that the transitions $B^*\bar{B}^*\to B\bar{B}^{*}$ and $B^*\bar{B}^*\to B\bar{B}$ do not vanish already in the pionless theory, such that
the coupled-channel dynamics for the states $Z_b'$ and $W_{b0}'$ produce an additional shift of the binding momenta and of the imaginary part
of the order $O(\gamma^2/\sqrt{m \delta})$~\cite{Baru:2016iwj}. Moreover, as it will be shown in this paper, once the
OPE interaction is included, the state $W_{b2}$ can also participate in the coupled-channel transitions, the latter having a significant impact on the
location of this state.

\section{One-pion exchange interaction}
\label{sec:OPE}

We start from the lowest-order nonrelativistic interaction Lagrangian~\cite{Hu:2005gf,Fleming:2007rp},
\begin{eqnarray}
{\cal L}=\frac{g_b}{{2}f_\pi}\left({\bm B^*}^\dagger \cdot {\bm\nabla}{\pi^a}\tau^a B
+B^\dagger\tau^a{\bm\nabla}{\pi^a} \cdot {\bm B^*} +i [{\bm B^*}^\dagger \times {\bm B^*}] \cdot {\bm\nabla}{\pi^a}\tau^a \right),
\label{lag}
\end{eqnarray}
and employ the heavy-quark flavour
symmetry to equate the dimensionless coupling constant $g_b$ to the similar constant $g_c$ which parametrises the
$D^*D\pi$ vertex. It can be determined directly from the observable $D^*\to D\pi$ decay width via
\begin{equation}
\varGamma(D^{*+}\to D^+\pi^0)=\frac{g_c^2m_{D^+} q^3}{24\pi f_\pi^2m_{D^{*+}}},
\label{D*width2}
\end{equation}
where $q$ is the centre-of-mass momentum in the final state. The numerical value extracted from data~\cite{Olive:2016xmw} 
reads
\begin{equation}
g_b=g_c\approx 0.57
\label{gc}
\end{equation}
and it agrees within 10\% with the recent lattice QCD determination of the $B^*B\pi$ coupling constant \cite{Bernardoni:2014kla}.

Depending on the channel, the OPE potential contains the pion propagator and two vertices of the type $B^*\to B^{(*)}\pi$
following directly from Lagrangian (\ref{lag}),
\begin{eqnarray}
v^a(B^*\to B\pi)&=&\frac{g_b}{2 f_\pi}\tau^a(\veep\cdot\veq),\nonumber \\[-2mm]
\label{vBBpi}\\[-2mm]
v^a(B^*\to B^*\pi)&=&-\frac{g_b}{\sqrt2 f_\pi}\tau^a(\veA \cdot\veq),\nonumber
\end{eqnarray}
where $\veA=\frac{i}{\sqrt{2}}(\veep\times {\veep'}^*)$, ${\veep}$ and ${\veep'}^*$ stand for the polarisation vectors of the initial and
final $B^*$ mesons, $\veq$ is the pion momentum, $\tau^a$ is the isospin Pauli matrix, and $f_{\pi}=92.2$ MeV is the pion decay
constant --- see also Refs.~\cite{Nieves:2011vw,Baru:2016iwj}.
Then, the OPE potentials connecting the heavy-meson $B^{(*)}B^{(*)}$ pairs in the initial and in the final state read\footnote{
Note that we use a convention where the integration weight is absorbed into the potential. Accordingly,
the usual factor $(2\pi)^3$ does not appear in the integral equation (\ref{Eq:JPC}) given below.}
\begin{eqnarray}
V_{B\bar B^*\to B^*\bar{B}}(\vep,\vep')&=&-\frac{2g_b^2}{(4\pi f_\pi)^2}\vetau_1\cdot\vetau_2^c\ {({\veep_1}\cdot \veq)({\veep_2'}^*\cdot \veq)}
\left(\frac{1}{D_{BB\pi}(\vep,\vep')}+\frac{1}{D_{B^*B^*\pi}(\vep,\vep')}\right),\nonumber \\
V_{B\bar B^*\to B^*\bar B^*}(\vep,\vep')&=& \frac{2\sqrt2 g_b^2}{(4\pi f_\pi)^2}\vetau_1\cdot\vetau_2^c\ {({\veA_1}\cdot \veq)({\veep_2}\cdot \veq)}
\left(\frac{1}{D_{BB^*\pi}(\vep,\vep')}+\frac{1}{D_{B^*B^*\pi}(\vep,\vep')}\right),\nonumber\\[0mm]
 \label{VBBpi}\\[-4mm]
 V_{B\bar B\to B^*\bar B^*}(\vep,\vep')&=&- \frac{2g_b^2}{(4\pi f_\pi)^2} \vetau_1\cdot\vetau_2^c\ {({\veep_1'}^*\cdot \veq)({\veep_2'}^*\cdot \veq)}
\left(\frac{2}{D_{BB^*\pi}(\vep,\vep')}\right),\nonumber\\
 V_{B^*\bar B^*\to B^*\bar B^*}(\vep,\vep')&=&-\frac{4 g_b^2}{(4\pi f_\pi)^2}\vetau_1\cdot\vetau_2^c\ {({\veA_1}\cdot \veq)({\veA_2}\cdot \veq)}
 \left(\frac{2}{D_{B^*B^*\pi}(\vep,\vep')}\right),\nonumber
\end{eqnarray}
where $\vep$ ($\vep'$) denotes the centre-of-mass momentum of the initial (final) heavy-meson pair and the pion momentum is
$\veq= \vep+\vep'$.
For the isovector states, the isospin factor which appears from the operator $\vetau_1\cdot\vetau_2^c$ is
\begin{equation}
\vetau_1\cdot\vetau_2^c=[3-2I(I+1)]_{I=1}=-1,
\label{isospincoeff}
\end{equation}
where $\vetau^c$ stands for the generator of the antifundamental representation, $\vetau^c=\tau_2\vetau^T\tau_2=-\vetau$.
Note that in contrast to what is given above, in the isoscalar channel, studied, for example, in Ref.~\cite{Baru:2016iwj},
the isospin factor is +3.

Furthermore, the expressions for the propagators in Eq.~(\ref{VBBpi}) read
\begin{eqnarray}\label{V13D1NR1}
D_{BB^*\pi}(\vep,\vep')&=&2E_{\pi}(\veq)\Bigl(m+m_{*}+\frac{\vep^2}{2m}+\frac{\vep'^2}{2m_{*}}+E_{\pi}(\veq)-\sqrt{s}\Bigr),\\
D_{BB\pi}(\vep,\vep')&=&2E_{\pi}(\veq)\Bigl(m+m+\frac{\vep^2}{2m}+\frac{\vep'^2}{2m}+E_{\pi}(\veq)-\sqrt{s}\Bigr),
\label{V22D1NR1}\\
D_{B^*B^*\pi}(\vep,\vep')&=&2E_{\pi}(\veq)\Bigl(m_{*}+m_{*}+\frac{\vep^2}{2m_{*}}+\frac{\vep'^2}{2m_{*}}+E_{\pi}(\veq)-\sqrt{s}\Bigr),
\label{V22D2NR1}
\end{eqnarray}
where $E_{\pi}=\sqrt{\veq^2+m_\pi^2}$ is the pion energy and $\sqrt{s}$ defines the total energy of the system which is conveniently represented
as $\sqrt{s}=m+m_*+E$, with $E$ counted with respect to the $B\bar{B}^*$ threshold.
The time-reversed transition potentials are trivially obtained from Eqs.~(\ref{VBBpi}) by interchanging $\vep$ and $\vep'$ and using that $D_{B^*B\pi}(\vep,\vep')=D_{BB^*\pi}(\vep',\vep)$.

The three-body effects are incorporated in the above OPE potentials via the heavy-meson recoil corrections and via the energy-dependent
terms in the time-ordered propagators from Eqs.~(\ref{V13D1NR1})-(\ref{V22D2NR1}).
The leading-order static OPE potentials are then easily obtained from the full results by neglecting the three-body terms, that is
by the replacement
\begin{equation}
\frac{1}{D_{B^{(*)}B^{(*)}\pi}(\vep,\vep')}+\frac{1}{D_{B^{(*)}B^{(*)}\pi}(\vep,\vep')} \to 2\times\frac{1}{2E_{\pi}^2(\veq)}
=\frac{1}{\veq^2+m_\pi^2}.
\end{equation}
We checked numerically that, for the mass spectra which are the focus of this work, the three-body terms
are basically irrelevant. Accordingly, one could as well have used the static pion propagators.
The reason is that the $B^{*}$-$B$ mass difference $\delta$ --- see Eq.~(\ref{deldef}) --- is about three times smaller than the pion mass and,
therefore, there are no three-body cuts located in the vicinity of $B^{(*)}B^{(*)}$ thresholds.
This situation is, however, very different in the $c$-sector:
the $D\bar D\pi$ threshold lies about 7 MeV below the $D\bar D^{*}$ threshold and,
hence, three-body effects play a prominent role for understanding the dynamics of the $X(3872)$ state treated as a $D\bar D^*$
molecule~\cite{Baru:2011rs,Baru:2013rta,Baru:2015tfa}.

The OEE can be calculated straightforwardly from the expressions given above if one makes the following replacements:
(i) the isospin factor of Eq.~(\ref{isospincoeff}) by +1; (ii) the pion mass by the $\eta$ mass, and (iii) the pion coupling constant
$g_c$ by the $\eta$ coupling constant
$g_c/\sqrt{3}$ which can be found, for example, in Ref.~\cite{Grinstein:1992qt}.

As soon as the OPE potential is included, it enables transitions to $D$ and even $G$ waves
which, therefore, have to be included in the extended set of basis states,
\begin{eqnarray}\label{Eq:basis}
&&0^{++}:\quad\{B\bar{B}({}^1S_0),B^*\bar{B}^*({}^1S_0),B^*\bar{B}^*({}^5D_0)\},\nonumber\\
&&1^{+-}:\quad\{B\bar{B}^*({}^3S_1,-),B\bar{B}^*({}^3D_1,-),B^*\bar{B}^*({}^3S_1),B^*\bar{B}^*({}^3D_1)\},\nonumber\\[
-3mm]
\label{basisvec}\\[-3mm]
&&1^{++}:\quad\{B\bar{B}^*({}^3S_1,+),B\bar{B}^*({}^3D_1,+),B^*\bar{B}^*({}^5D_1)\},\nonumber\\
&&2^{++}:\quad\{B\bar{B}({}^1D_2),B\bar{B}^*({}^3D_2),B^*\bar{B}^*({}^5S_2),B^*\bar{B}^*({}^1D_2),B^*\bar{B}^*({}^5D_2),
B^*\bar{B}^*({}^5G_2)\}.\nonumber
\end{eqnarray}
As before, the C-parity of the states is indicated explicitly in parentheses whenever necessary.

The coupled-channel integral equations for the scattering amplitude take the form
\begin{eqnarray}
a^{(JPC)}_{ij}(p,p')=V^{(JPC)}_{ij}(p,p')-\sum_n \int dk\,k^2 V^{(JPC)}_{in}(p,k){G_n(k)}a^{(JPC)}_{nj}(k,p'),
\label{Eq:JPC}
\end{eqnarray}
where $i,j$, and $n$ label the basis vectors, as they appear in Eq.~(\ref{Eq:basis}), and
\begin{equation}
G_n=\left(k^2/(2\mu_n)+m_{1,n}+m_{2,n}-\sqrt{s}-i\epsilon\right)^{-1},\quad \mu_n=\frac{m_{1,n}m_{2,n}}{m_{1,n}+m_{2,n}}
\end{equation}
are the propagator and the reduced mass of the heavy
meson-antimeson pair in the given channel evaluated in its centre of mass.
The potential $V^{(JPC)}_{ij}(p,p')$ includes the contact term as well as the meson exchange interactions in the channel
with the given quantum numbers $J^{PC}$ and in the given partial wave.
The Lippmann-Schwinger-type equations derived are regularised with the sharp cutoff, where the value of the latter
is chosen to be of the order of a natural
hard scale in the problem --- see Ref.~\cite{Baru:2016iwj}.
All necessary details of the coupled-channel formalism, of the
partial wave projection, and so on can be
found in Refs.~\cite{Baru:2016iwj,Baru:2017pvh}. However, it is important to pinpoint a crucial difference between the potentials used in
Refs.~\cite{Baru:2016iwj} and the ones from this work: while the OPE potential from Ref.~\cite{Baru:2016iwj} describes the interaction in the isosinglet
channels, here it operates in the isovector systems. Thus, the trivial modification one needs to make in the equations is to change the isospin
coefficient 3, which multiplies the OPE potential in the isosinglet channels, for the isospin coefficient -1, which corresponds to the isovector channels --- see
Eq.~(\ref{isospincoeff}). In addition, it has to be noticed that the sign of the OPE potential also depends on the C-parity of the system.
Then, as a net result, the central ($S$-wave) part of the OPE potential is attractive in the C-odd $Z_b$'s channel
while it is repulsive in the C-even channels where the
$W_{bJ}$ states are predicted. It is important to note, however, that the net effect of the OPE can be
attractive also in those channels due to the nondiagonal transition potentials --- in particular, due to the $S$-$D$ transitions
induced by the tensor force.

The inclusion of the OPE interaction modifies the potentials (\ref{Vct0++})-(\ref{Vct2++}). Nevertheless, as demonstrated in
Ref.~\cite{Baru:2016iwj}, in the strict heavy-quark limit, if all coupled channels and all relevant partial waves are taken into account, for a
given set of quantum numbers $J^{PC}$, one can find the unitary operator that block diagonalises
the potential into the form
(for the sake of transparency, the size of the blocks is quoted explicitly in parentheses),
\begin{eqnarray}
&&\tilde V^{(0++)}(3\times 3)=A(2\times 2)\oplus B(1\times 1),\nonumber\\
&&\tilde V^{(1+-)}(4\times 4)=A(2\times 2)\oplus B(1\times 1)\oplus C(1\times 1),\nonumber\\[-3mm]
\label{Vtild}\\[-3mm]
&&\tilde V^{(1++)}(3\times 3)=A(2\times 2)\oplus D(1\times 1),\nonumber\\
&&\tilde V^{(2++)}(6\times 6)=A(2\times 2)\oplus D(1\times 1)\oplus E(3\times 3).\nonumber
\end{eqnarray}
This conclusion does not depend on the isospin of the system and it is, therefore, equally valid for both isosinglets and isovectors.
Since the contact interactions contribute to both matrix $A$ (via $C_1=C_{1a}+C_{1b}$) and matrix $B$ (via ${C_1^\prime}=C_{1a}-3C_{1b}$) and since
at least one of these two matrices enters all four
potentials for all quantum numbers one expects that the existence of the $Z_b$ and $Z_b'$ as bound states in
the $1^{+-}$ channel entails the existence of the bound states in the other three channels, too, with the degeneracy pattern
given by Eq.~(\ref{HQSSlimit}) as in case of the purely contact theory.
As explained in the previous section, this pattern is expected to be lifted
as soon as the physical masses of the $B$ and $B^*$ mesons are used in the calculations. Meanwhile, the HQSS constraints are
still expected to work quite well for the contact potentials.

\section{Results and discussion}
\label{sec:results}

In this section, we present and discuss the results of our calculations with special emphasis
on the role of the pion dynamics for the
location of the poles of the $W_{bJ}$'s --- the spin partners of the $Z_b$ states. Furthermore, motivated by the fact that in the isovector
channel the relative importance of the $\eta$-exchange is larger than in the isoscalar channel, we also include it explicitly in
our calculations.

As our starting point,
we assume both $Z_b$ and $Z_b'$ to be shallow bound states and treat their binding energies as input parameters.
This allows us to fix completely the HQSS-constrained leading-order contact potential --- see
Eqs.~(\ref{Vct0++})-(\ref{Vct2++}).
For definiteness, as an example, we fix the binding energies of the $Z_b$'s to be
\begin{equation}
E_B(B^*\bar B)[Z_b]=5~\mbox{MeV},\quad E_B(B^*\bar B^*)[Z_b']=1~\mbox{MeV},
\label{input1}
\end{equation}
in line with Ref.~\cite{Cleven:2011gp}. {For both $Z_b$'s the energy is counted relative to the respective reference threshold explicitly stated 
in Eq.~(\ref{input1}) in parentheses.}
Ideally, one would need to extract pole locations and residues  for the spin partner states of the $Z_b$'s directly from the
calculated  $T$-matrices. However,  this requires an analytic continuation of the amplitudes into the complex plain 
that goes beyond the scope of the present paper. Therefore, to extract the  binding energies and the widths of the spin partners, $W_{bJ}$'s, 
we mimic the
experimental procedure and calculate the production rates {in the elastic channels} for the spin partner states.
In what follows, similarly to Eq.~(\ref{input1}), the binding energies of the partner states will be always defined relative
to their reference thresholds, namely,
{\be
\begin{array}{ccl}
\mbox{Threshold:}&\hspace*{5mm}&\mbox{Spin partner:}\\
B\bar{B}&&~W_{b0}(0^{++}), \\
B\bar{B}^*&&~W_{b1} (1^{++}),\\
B^*\bar{B}^*&&~W'_{b0}(0^{++}),~W_{b2}(2^{++}).
\end{array}
\ee}

\subsection{Dependence on the HQSS breaking scale}

\begin{figure}
\includegraphics[width=0.5\textwidth]{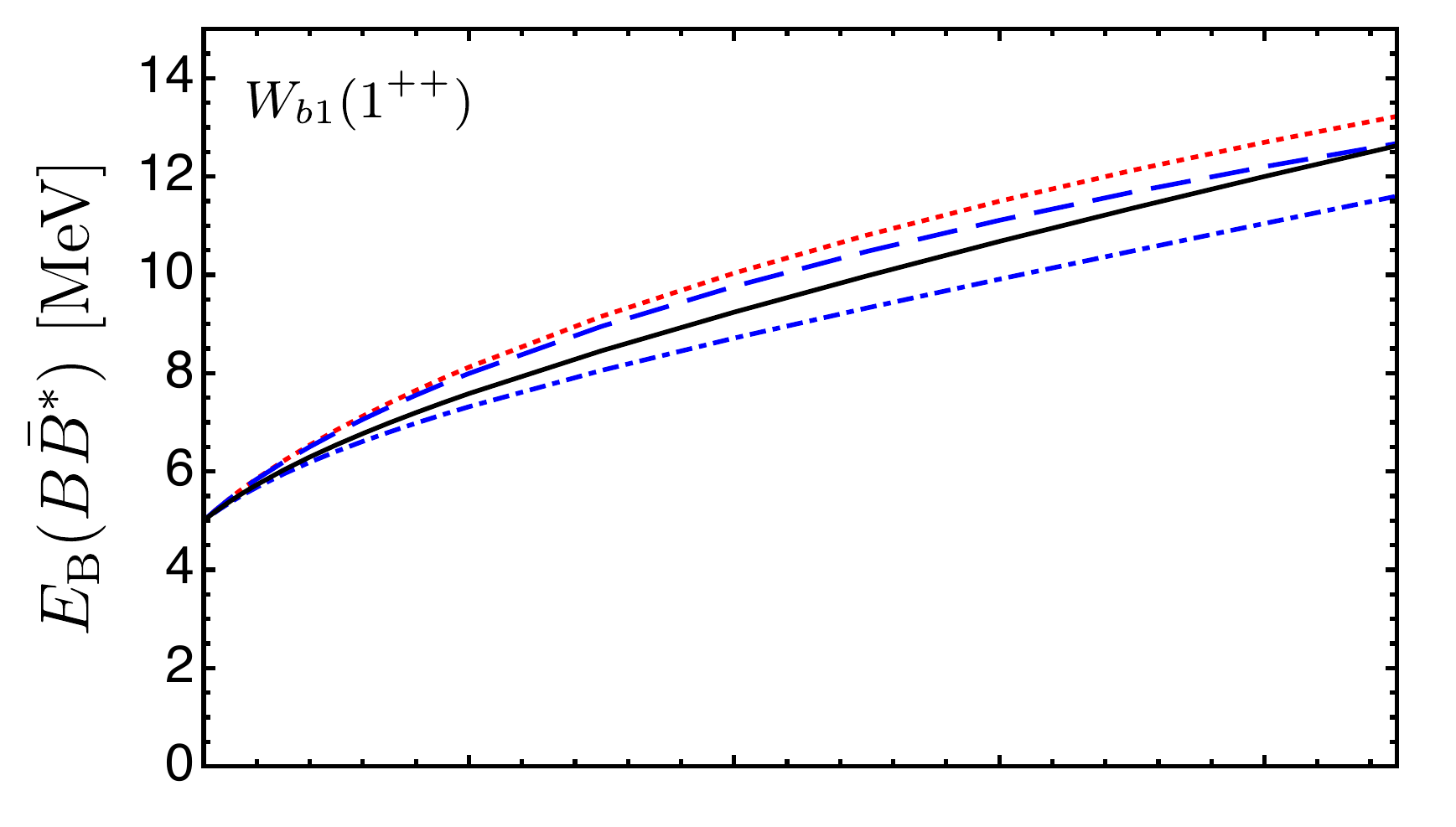}
\includegraphics[width=0.5\textwidth]{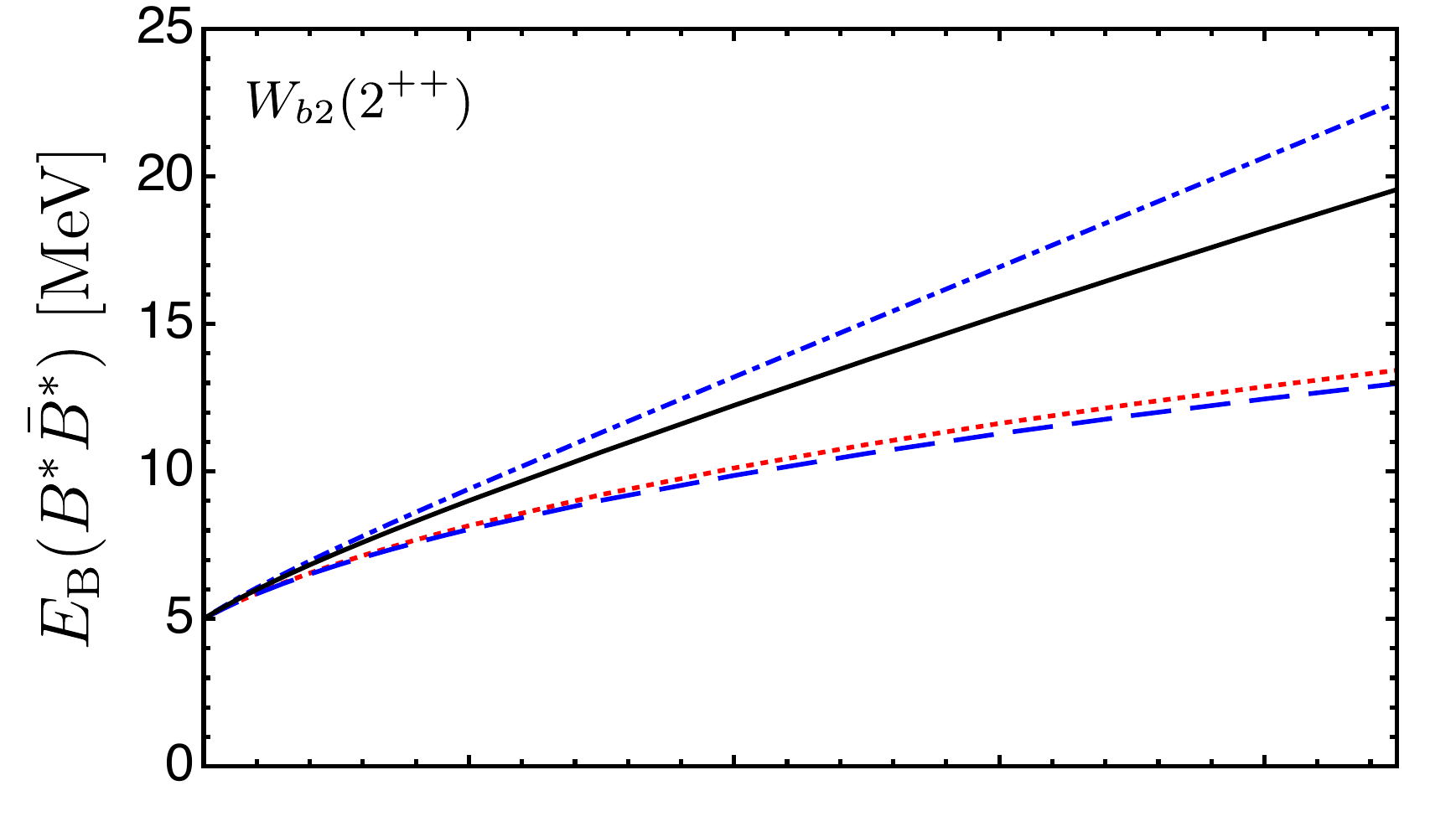}\\
\includegraphics[width=0.5\textwidth]{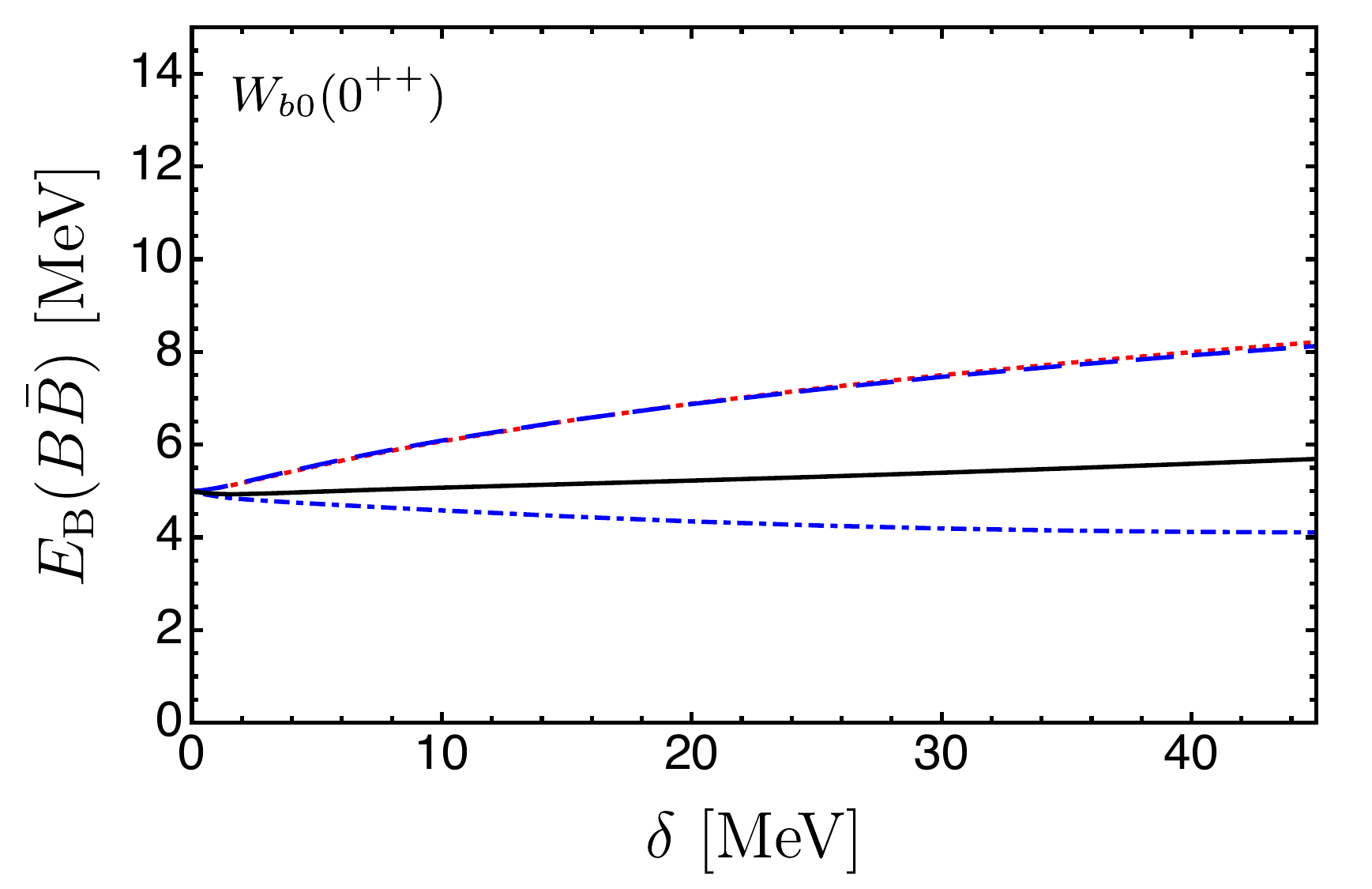}
\includegraphics[width=0.5\textwidth]{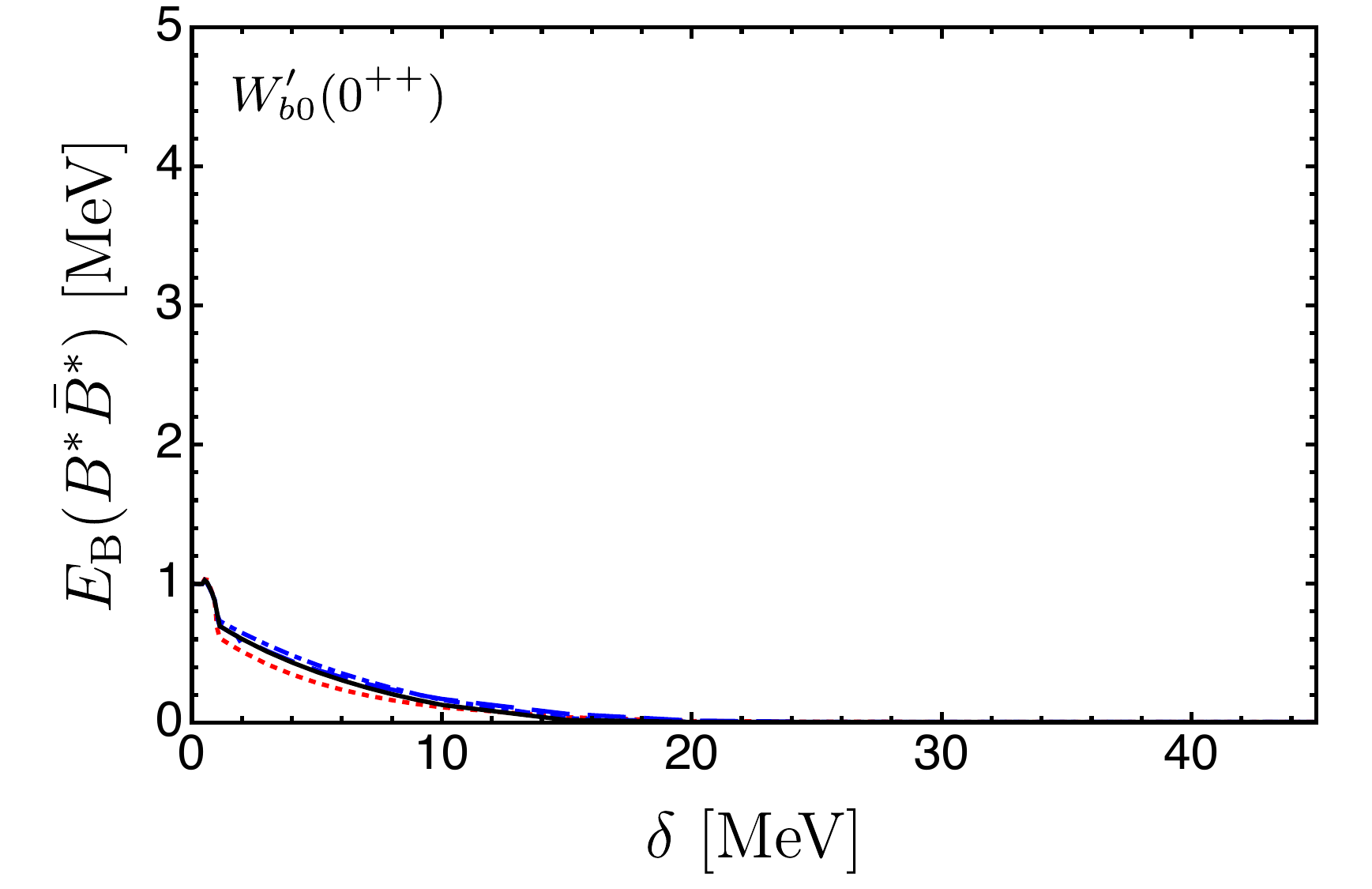}
\caption{Evolution of the binding energies of the $Z_b$'s spin partners calculated with and without $\pi$- and $\eta$-meson
exchanges as functions of the mass splitting $\delta$ between the $B^*$ and $B$ mesons. The contact terms are re-fitted for each value of $\delta$
to provide the given binding energies of the $Z_b$ and $Z_b'$ states used as input --- see Eq.~(\ref{input1}). The binding energies of the $W_{bJ}$ states are defined relative
to their reference thresholds quoted in parentheses.
The red dotted curves correspond to the pionless (purely contact) theory {--- Scenario A}; 
the blue dashed curves are obtained for the central ($S$-wave) part of the
OPE included {--- Scenario B}; the blue dashed-dotted lines represent the results for the full OPE, including tensor 
forces {--- Scenario C};
the black solid curves show the results of the full calculation with both OPE and OEE included on top of the contact 
interactions {--- Scenario D}.
The physical limit corresponds to the right edge of the plots. The results are obtained with the sharp cutoff $\Lambda =1$ GeV
in the integral equations (\ref{Eq:JPC}). The uncertainty caused by the residual $\Lambda$-dependence of the equations
can be estimated using the results presented in Table~\ref{tab:res}.
}
\label{fig:EBvsDelta}
\end{figure}

To highlight the impact of the OPE and OEE on the pole locations of the spin partner states,
in Fig.~\ref{fig:EBvsDelta}, we show the evolution of the extracted binding energies as functions of
$\delta$ --- the HQSS breaking mass splitting between the $B$ and $B^*$ mesons --- for four different scenarios:
\begin{itemize}
\item {Scenario A:} purely contact potential;
\item {Scenario B:} contact potential plus the central ($S$-wave) OPE interaction;
\item {Scenario C:} contact potential plus the full OPE interaction;
\item {Scenario D:} contact potential plus the full OPE and OEE interactions altogether.
\end{itemize}

It follows from Eqs.~(\ref{HQSSlimit}) and (\ref{Vtild}) that, in the strict HQSS limit ($\delta=0$), the $Z_b$'s and the $W_{bJ}$'s
populate two families of states {and that this conclusion holds} both with and without OPE and OEE interactions included. 
{Given the input from Eq.~(\ref{input1}),} this explains why, for all four scenarios above,
the $2^{++}$, $1^{++}$, and one of the $0^{++}$
states have the same binding energy $E_B=5$~MeV while the other
$0^{++}$ spin partner has the binding energy $E_B=1$~MeV in this limit.
{However, for $\delta>0$, HQSS is lifted and the exact degeneracy (\ref{HQSSlimit}) is lost --- see the discussion
in Secs.~\ref{sec:contact} and~\ref{sec:OPE} above. Then, as $\delta$ grows from zero to the physical value of 45~MeV,
the spin partners from the first family ($W_{b0}$, $W_{b1}$, and $W_{b2}$) tend to become more bound while the state $W_{b0}'$
becomes unbound fast and its pole moves to the second Riemann
sheet.}\footnote{Strictly speaking, the $0^{++}$ state residing near the $B^*\bar{B}^*$ threshold
should be regarded as a resonance since
it appears above the $B\bar{B}$ and $B\bar{B}^*$ thresholds and the corresponding
pole acquires an imaginary part. Nevertheless, for clarity, we still
refer to it as to a bound or virtual state with respect to the $B^*\bar{B}^*$ threshold and use the language of the corresponding two-sheet
Riemann surface.} Note that while varying $\delta$ we keep adjusting the parameters of the contact
potential such that the pole locations of $Z_b$ and $Z_b'$ stay fixed at the values given in Eqs.~(\ref{input1}).

From Fig.~\ref{fig:EBvsDelta} one can conclude that the central ($S$-wave) part of the OPE has no influence on the position of the
poles\footnote{More rigorously, the $S$-wave part of the OPE is absorbed into the contact interactions via the re-fit of the latter to preserve the
location of the $Z_b$ poles.} while
its tensor part has a significant impact, especially on the $2^{++}$ state $W_{b2}$. {Indeed,} due to the additional attraction 
which stems from the OPE-related
tensor forces, the binding energy of the $W_{b2}$ grows by almost a factor of two as compared to the results of the
pionless theory (from 13 to 23 MeV).
Furthermore, by comparing the results in the strict HQSS limit ($\delta=0$) with those for the physical mass splitting 
{($\delta=45$~MeV)}, one can conclude
that spin symmetry is violated
quite strongly and that a substantial amount of this violation for the $2^{++}$ state stems from the tensor forces.

Finally, it can also be seen in Fig.~\ref{fig:EBvsDelta} that the $\eta$-exchange does not play a prominent role for the systems under study, for it
only slightly diminishes the OPE.  This can be seen  in all four plots
in Fig.~\ref{fig:EBvsDelta}.  

\subsection{{Dependence on the one-boson-exchange  strength parameter}}

\begin{figure}
\includegraphics[width=0.5\textwidth]{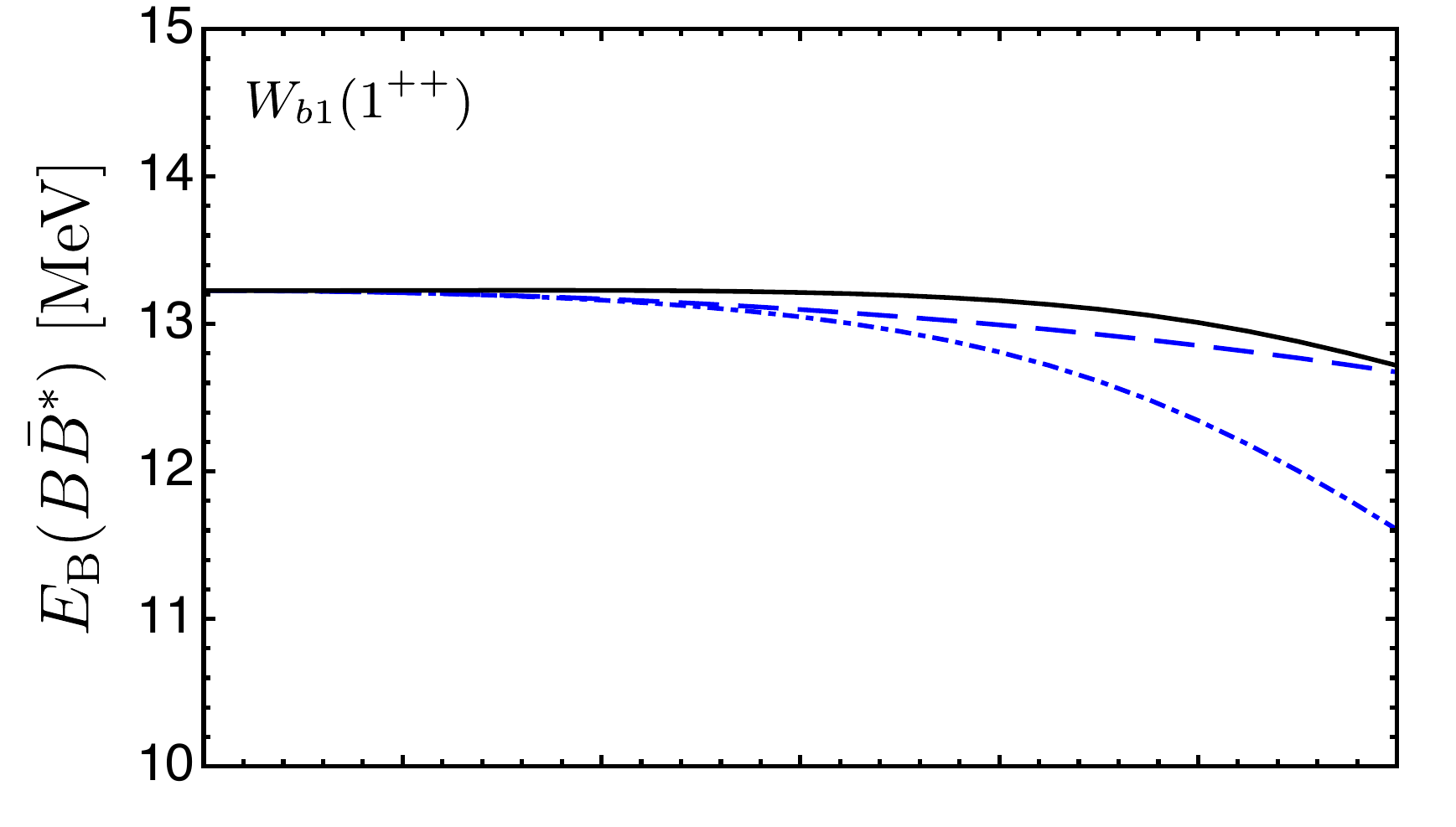}
\includegraphics[width=0.5\textwidth]{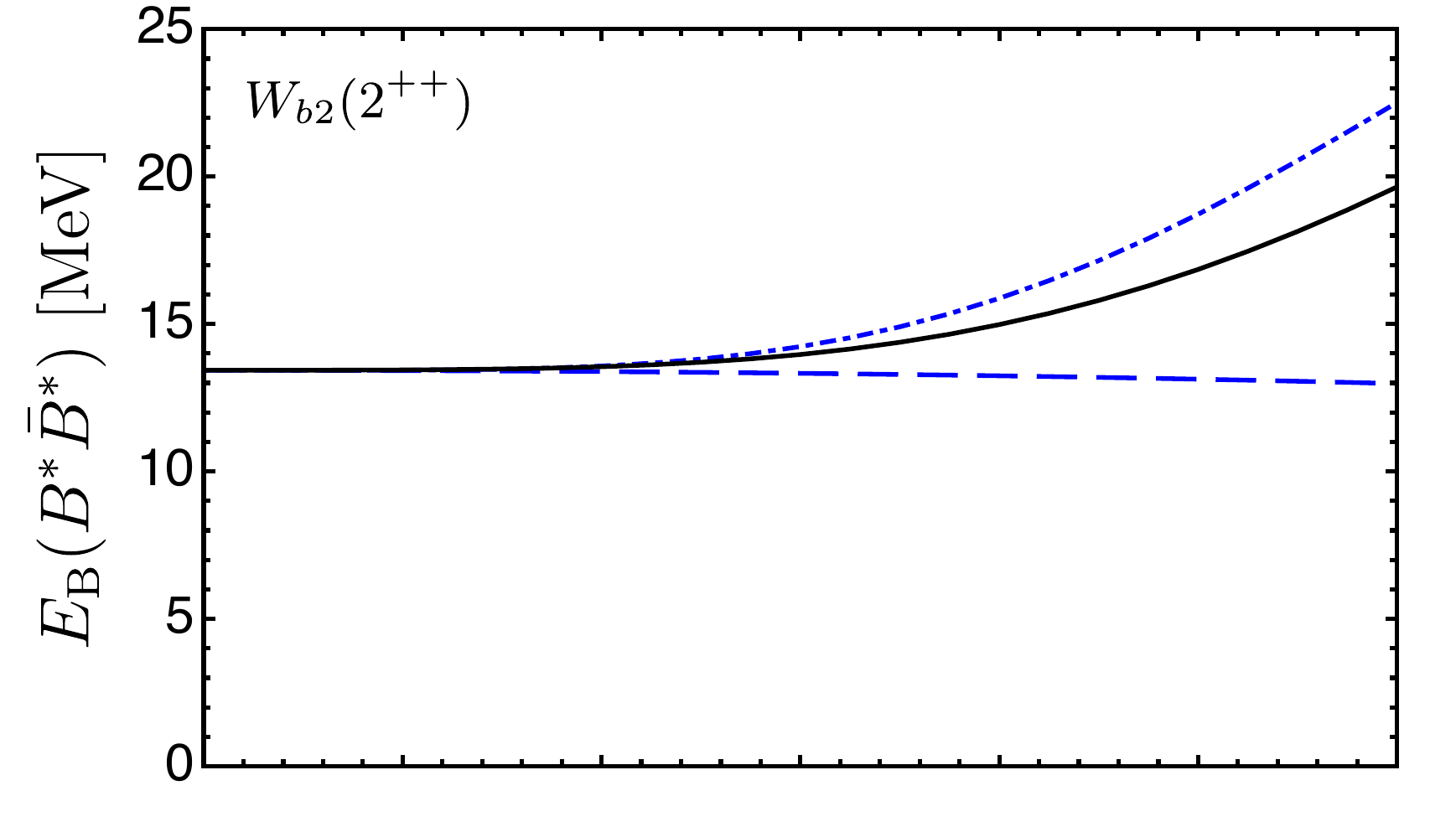}
\includegraphics[width=0.5\textwidth]{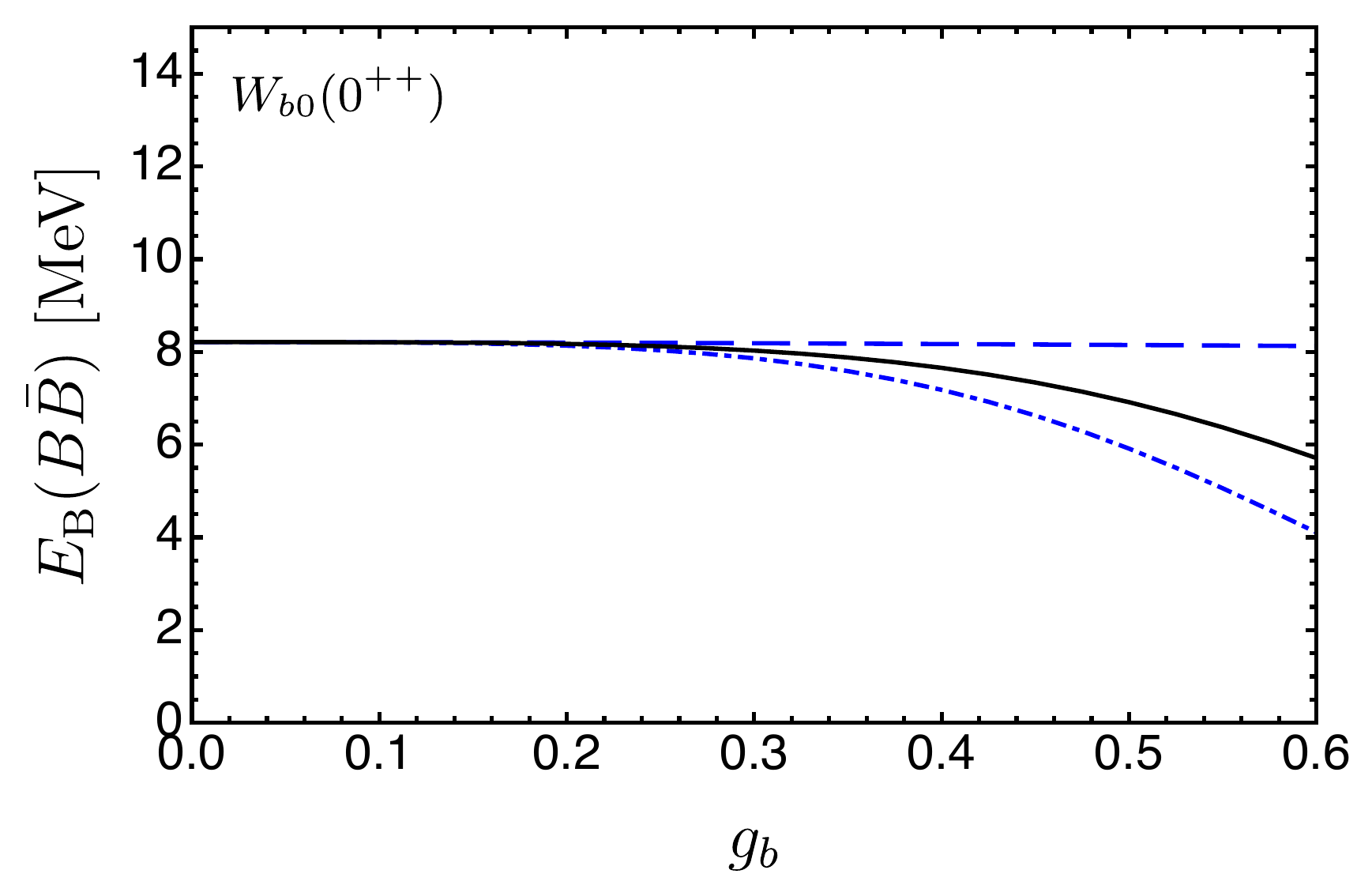}
\includegraphics[width=0.5\textwidth]{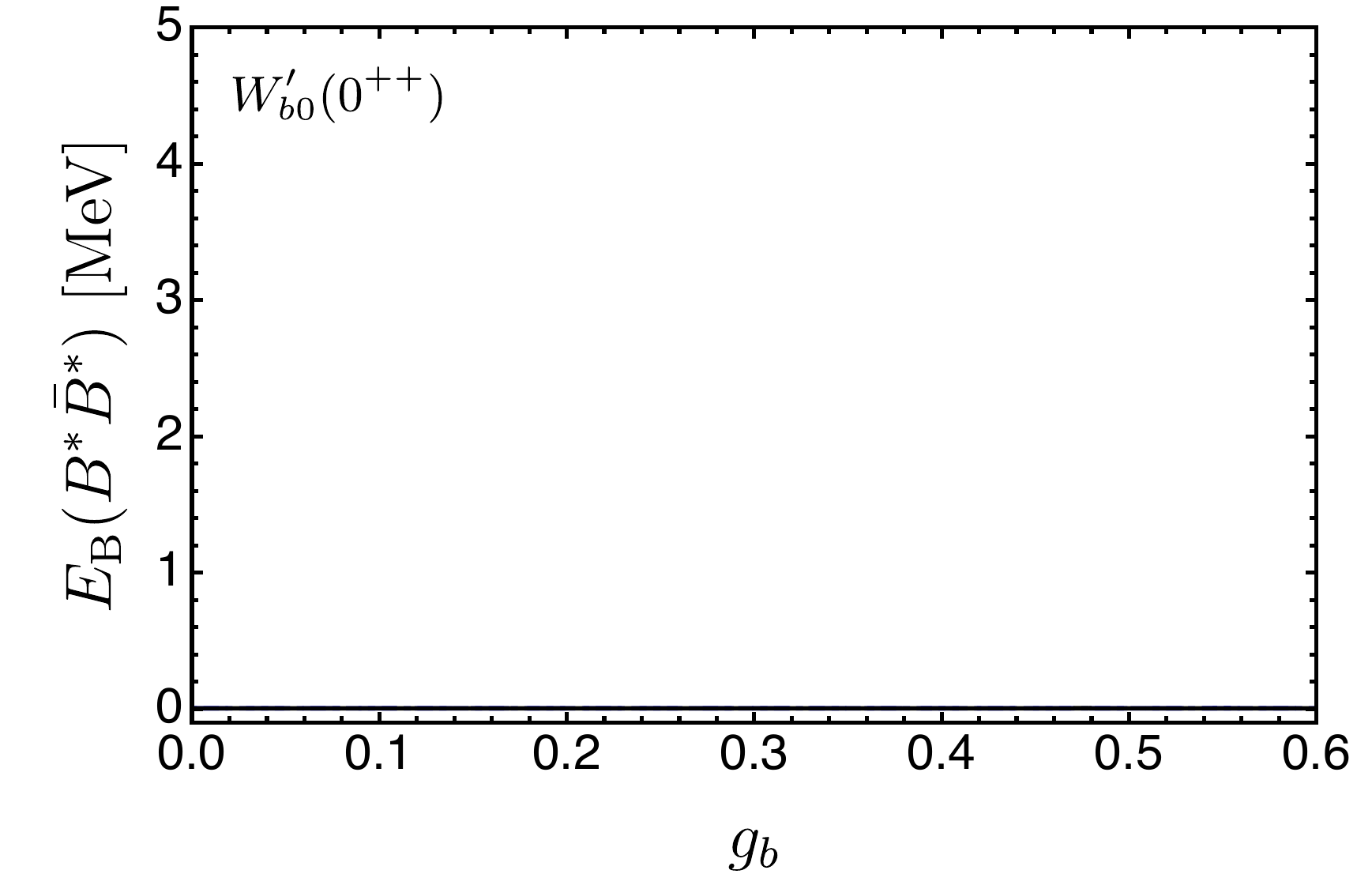}
\caption{The binding energies of the $Z_b$'s spin partners as functions of the coupling constant $g_b$ for the physical
mass splitting $\delta$ between $B^*$ and $B$ mesons. The binding energies of the $Z_b$ and $Z_b'$ states
are used as input --- see Eq.~(\ref{input1}).
The notation of curves is the same as in Fig.~\ref{fig:EBvsDelta}. The results are obtained with the sharp cutoff $\Lambda =1$ GeV in the integral equations
(\ref{Eq:JPC}).}
\label{fig:EBvsgB}
\end{figure}

To further illustrate the role played by the dynamics governed by the pion and $\eta$-meson exchanges, in Fig.~\ref{fig:EBvsgB},
we plot the variation of the binding energies for the spin partners $W_{bJ}$ with the coupling constant $g_b$ varied from 0 (the results
of the purely contact theory are naturally recovered in this limit) to its physical value
quoted in Eq.~(\ref{gc}). In this calculation, the masses of the $B$ and $B^*$ mesons are fixed to their physical values~\cite{Olive:2016xmw}.
One can draw several conclusions from Fig.~\ref{fig:EBvsgB}. On the one hand, for all values of $g_b$
the effect of the central $S$-wave OPE potential can always be absorbed completely into the contact terms. 
{Indeed, in all plots, the blue dashed line behaves like a constant}.
On the other hand, starting from $g_b\simeq 0.3$, {the dashed-dotted line starts to deviate from the dashed line indicating that}
the role of the tensor forces increases fast thus providing
a substantial shift in the binding energy in the physical limit for the $g_b$ --- this effect is best seen for the tensor partner $W_{b2}$.

\begin{figure}
\includegraphics[width=0.50\textwidth]{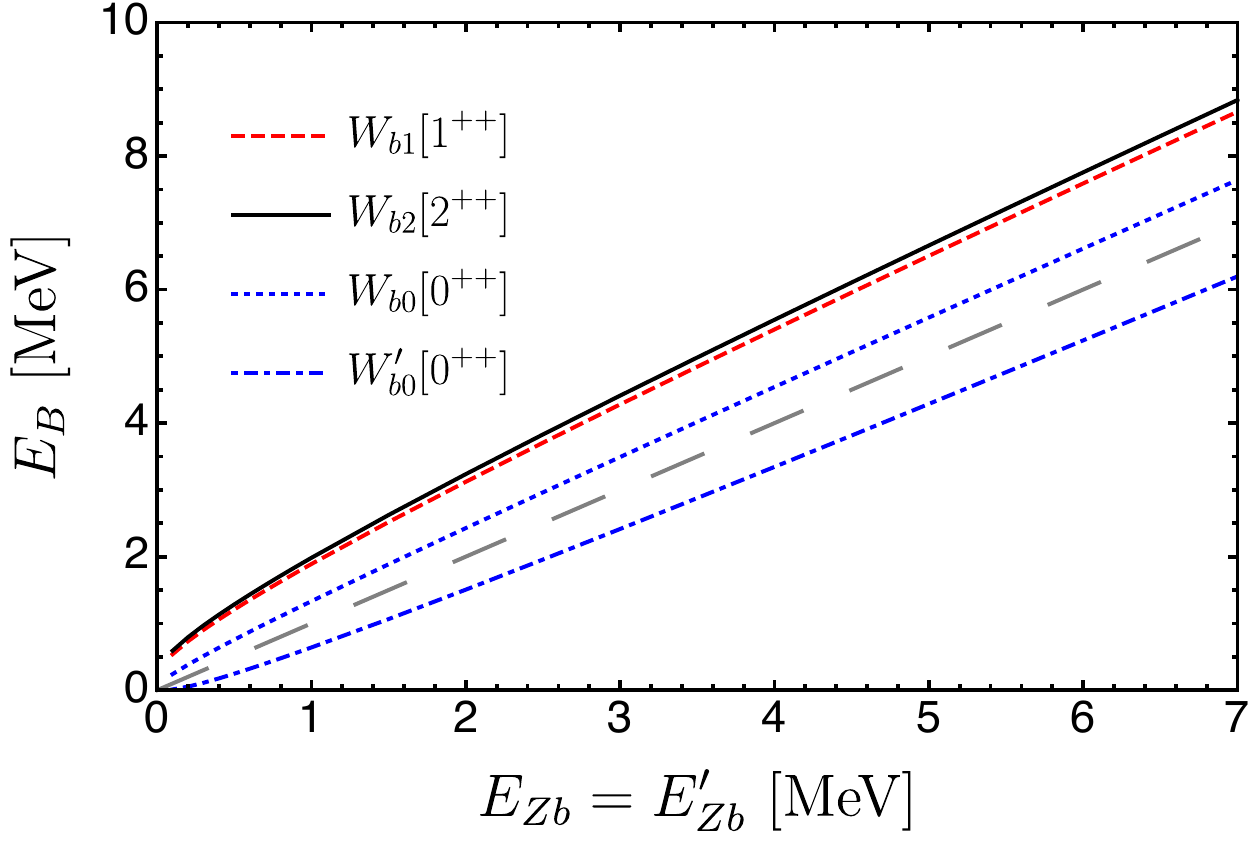}
\includegraphics[width=0.50\textwidth]{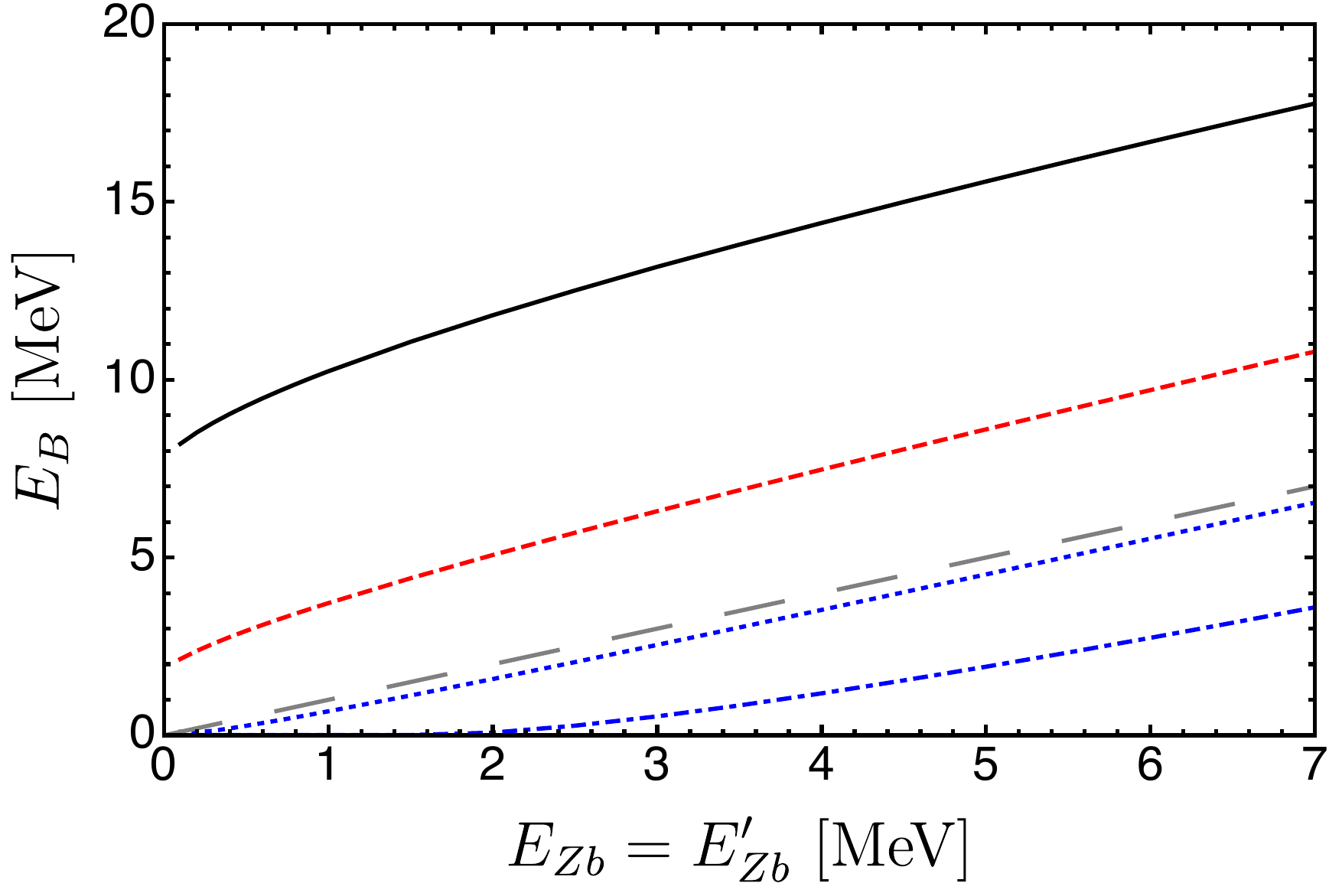}
\caption{Dependence of the binding energies of the $W_{bJ}$'s on the input used for the binding energies of the $Z_b$'s for the purely contact
interaction (the first plot) and
for the full theory, including the OPE and OEE (the second plot). To guide the eye, the (equal) binding energies of the $Z_b$'s are shown as the grey dashed lines.
The results are obtained with the sharp cutoff $\Lambda =1$ GeV in the integral equations (\ref{Eq:JPC}).}
\label{fig:EBWpionful}
\end{figure}

\begin{figure}
\centering
\includegraphics[width=0.508\textwidth]{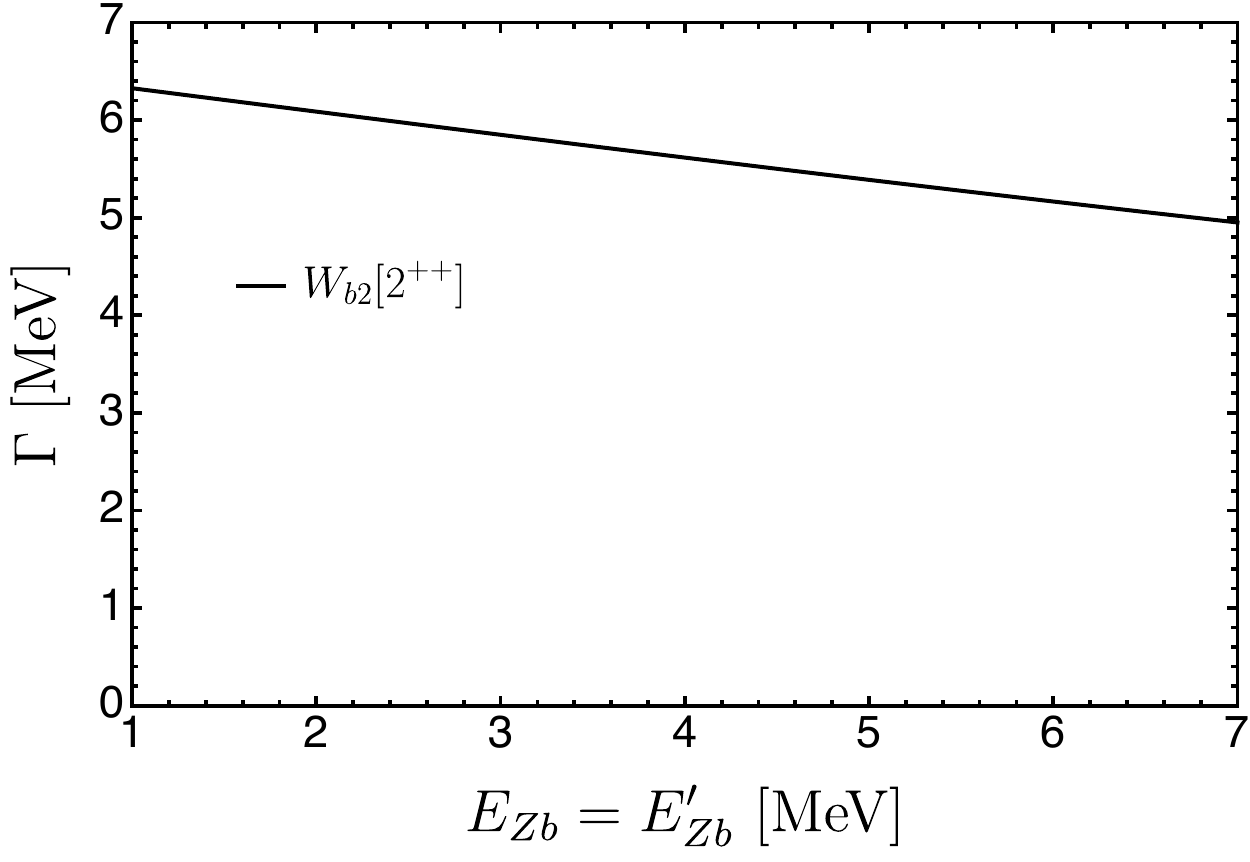}
\caption{ The decay width of the $W_{b2} (2^{++})$ state from coupled-channel transitions $B^*\bar B^*\to B\bar B$ and $B^*\bar B^*\to B\bar B^*$
evaluated in the full theory, including the OPE and OEE. \label{fig:GBWfull}}
\end{figure}

\subsection{Dependence on the input for the $Z_b$'s binding energies}

Finally, we investigate the dependence of the binding energies of the spin partners $W_{bJ}$'s on the input used for binding energies of the $Z_b$'s.
To simplify the presentation of the results, we choose coinciding binding energies of the latter and vary them
in a sufficiently wide range compatible with the values found in the literature. {The results for the binding energies 
of the spin partners $W_{bJ}$'s are given in Fig.~\ref{fig:EBWpionful}.
To guide the eye, the constraint $E_B[Z_b']=E_B[Z_b]$ is shown by the grey dashed line. Furthermore, we consider the limit
$E_B[Z_b']=E_B[Z_b]\to 0$ which provides a smooth matching between the case of the $Z_b$'s as bound states and as virtual levels.
The results for the widths of the partners are shown in Fig.~\ref{fig:GBWfull}.}
It is instructive to note that, in agreement with Eq.~(\ref{HQSSlimit}), in the strict HQSS limit and for $E_B[Z_b]=E_B[Z_b']$,
all six partner states are strictly
degenerate. Therefore, deviations of the curves for the spin partner states $W_{bJ}$'s from the grey dashed line in
Fig.~\ref{fig:EBWpionful} illustrate the importance of the HQSS breaking corrections related to
the nonvanishing $B^{*}$-$B$ mass splitting. In particular, had the strict HQSS limit been realised in nature all six spin partner states would have become
unbound in unison for $E_B[Z_b]=E_B[Z_b']\to 0$.
{Actually}, as it is seen from {the left plot in} Fig.~\ref{fig:EBWpionful}, 
a similar pattern is observed in the purely contact theory beyond the strict HQSS limit.
On the other hand, in the full theory with pions and $\eta$-mesons {(see the right plot in Fig.~\ref{fig:EBWpionful})},
the $1^{++}$ and $2^{++}$ partners survive as bound states in the limit $E_B[Z_b]=E_B[Z_b']\to 0$ when both $Z_b$'s turn to virtual levels.
The width of the $2^{++}$ state is expected at the level of a few MeV --- see Fig.~\ref{fig:GBWfull} --- so that one should be able to
resolve it from the $B^*\bar{B}^*$ threshold. Meanwhile, the poles for both $0^{++}$ partners move to the unphysical Riemann sheets.

We note that the inclusion of the OPE interaction leads to selfconsistent results only
if all relevant partial waves as well as particle channels which are coupled via the pion-exchange potential are taken into account. 
{Conversely,} the partial neglect of the coupled-channel dynamics
not only significantly affects the predictions for the partner states but also leads to the regulator(cutoff)-dependent results
already in the strict HQSS limit. This is in full agreement with the results in the $c$-quark sector reported in Ref.~\cite{Baru:2016iwj}.

In agreement with the discussion in Sec.~\ref{sec:contact}, the results of the full coupled-channel problem
beyond the HQSS limit may also exhibit some cutoff dependence which is however {quite mild and which} can be absorbed into a redefinition of higher-order
HQSS violating contact interactions. To illustrate that the effect of the cutoff variation on our results is minor, in Table~\ref{tab:res} we estimate the uncertainty in the
binding energies and widths of the partner states when the cutoff is varied in the range from 800 to 1500 MeV.

\begin{table}[h]
\begin{center}
\begin{tabular}{ | l | c |c |c |c |c | r |}
\hline
& $Z_{b1} (1^{+-})$ & $Z'_{b1} (1^{+-}) $ & $ W_{b0} (0^{++})$ & $W'_{b0} (0^{++})$ & $W_{b1} (1^{++})$ & $W_{b2} (2^{++})$
\\ \hline
$E_B$ \mbox{[MeV]} &5 \mbox{(input)} &1 \mbox{(input)} & 5.3 $\pm$1.7& --- & 12.4$\pm$0.6 & 19.8$\pm$2.2\\ \hline
$\Gamma_B$ \mbox{[MeV]} & --- & --- 
& --- & --- & --- & 4.6 $\pm$ 1.0\\ \hline\hline
$E_B$ \mbox{[MeV]} &1 \mbox{(input)} &1 \mbox{(input)} & 0.7 $\pm$0.5& --- & 3.8$\pm$0.1 & 10.2$\pm$1.8\\ \hline
$\Gamma_B$ \mbox{[MeV]} & --- & --- 
& --- & --- & --- & 6.2 $\pm$ 1.1\\ \hline\hline
\hline
\end{tabular}
\caption{
The binding energies and the widths of the spin partners for the two sets of the $Z_b$ and $Z_b'$ binding energies used as
input parameters. The uncertainty
in the results is due to the variation of the cutoff in the Lippmann-Schwinger equations from 800 to 1500 MeV. 
\label{tab:res} }
\end{center}
\end{table}
{Note that, in the vicinity of the peak, 
the energy dependence of the line shape for the
$2^{++}$ partner state has a clear Breit-Wigner form, from which the parameters quoted in Table~\ref{tab:res} 
were extracted. }
On the contrary, since the $Z_b'$ state resides very close to the $B^*\bar B^*$ threshold, the Breit-Wigner distribution in the resonance region 
is strongly distorted by threshold effects. Then, to arrive at the quantity that can be compared to the width from experiment one would need to convolute
the energy distribution from the coupled-channel amplitudes with the resolution function, integrate it over the energy bins and then analyse 
with the standard Breit-Wigner techniques. Since this goes beyond the scope of the current work, we do not quote the width of the $Z_b'$ in Table~\ref{tab:res}.
 
\section{Comment on the pionless theory}
\label{sec:comment}

In the previous section, we demonstrated that nonperturbative pion exchange has a significant impact on
the pole locations of the spin partner states for the $Z_b$ and $Z_b'$. In this section, we show that it plays
an additional important role to identify correctly the realistic solution from the pair of solutions
present in the purely contact theory as long as input from the $1^{+-}$ channel is used to fix the
parameters. Indeed, in the potential (\ref{Vct1+-}) written in terms of the low-energy constants $C_1$ and $C_1'$,
an interchange of the latter only changes the sign of the off-diagonal
elements which has no effect on the observables as long as Eq.~(\ref{Vct1+-})
is the only contribution to the potential. Therefore, in the purely contact theory there are two solutions
that both lead to the same masses of the $Z_b$ states.
Accordingly, there is an alternative solution to the one given in Eq.~(\ref{HQSSlimit}) for the spin partner states in the strict HQSS limit, namely
\begin{eqnarray}
\mbox{Sol.2}:~
E_B^{(0)}[Z_b]=E_B^{(0)}[W_{b0}]\quad\mbox{and}\quad E_B^{(0)}[Z_b']=E_B^{(0)}[W'_{b0}]=E_B^{(0)}[W_{b1}]=E_B^{(0)}[W_{b2}].
\label{Sol.2}
\end{eqnarray}

\begin{figure}
\hspace{-0.4cm}
\includegraphics[width=0.5\textwidth]{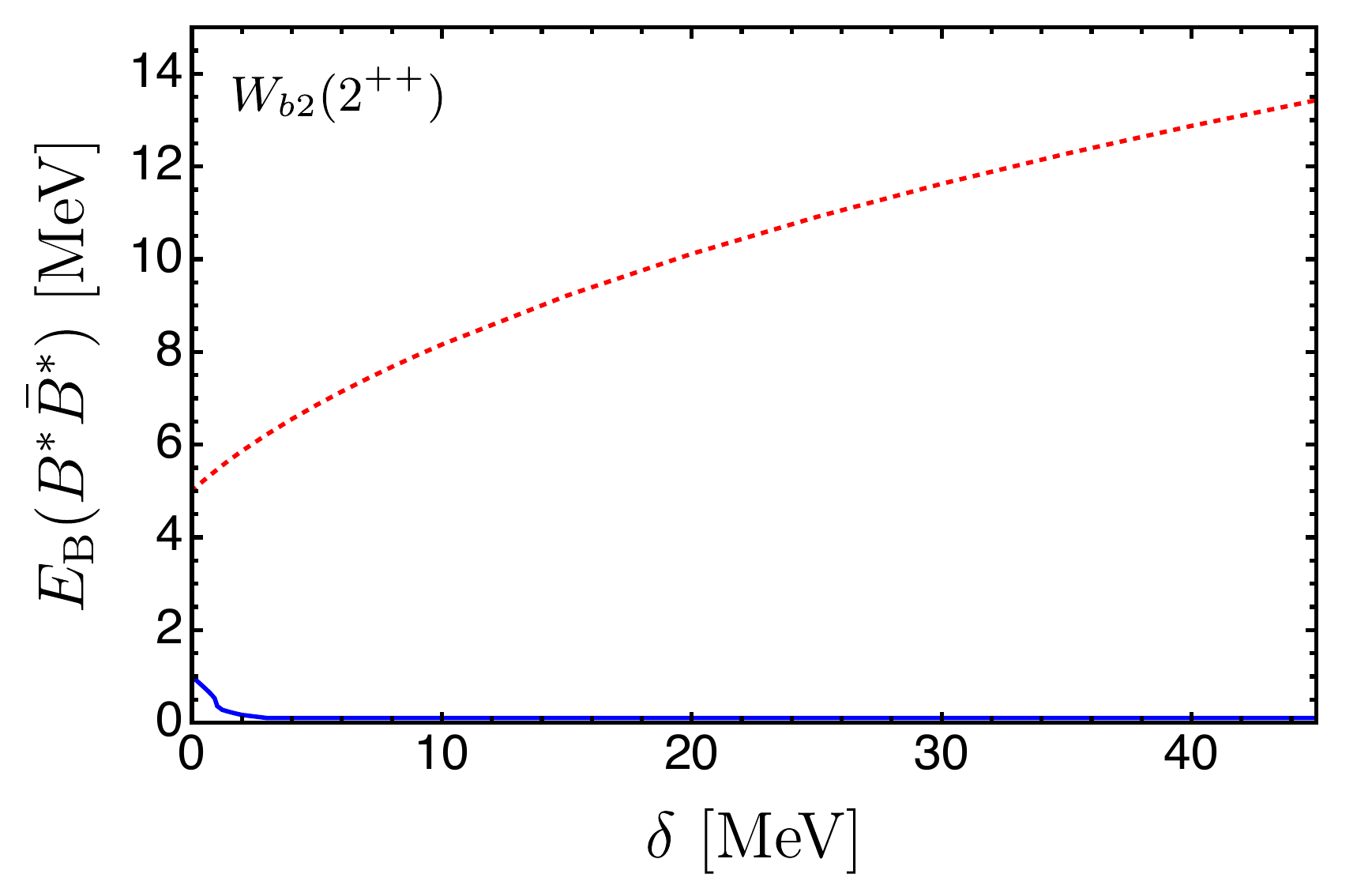}
\includegraphics[width=0.5\textwidth]{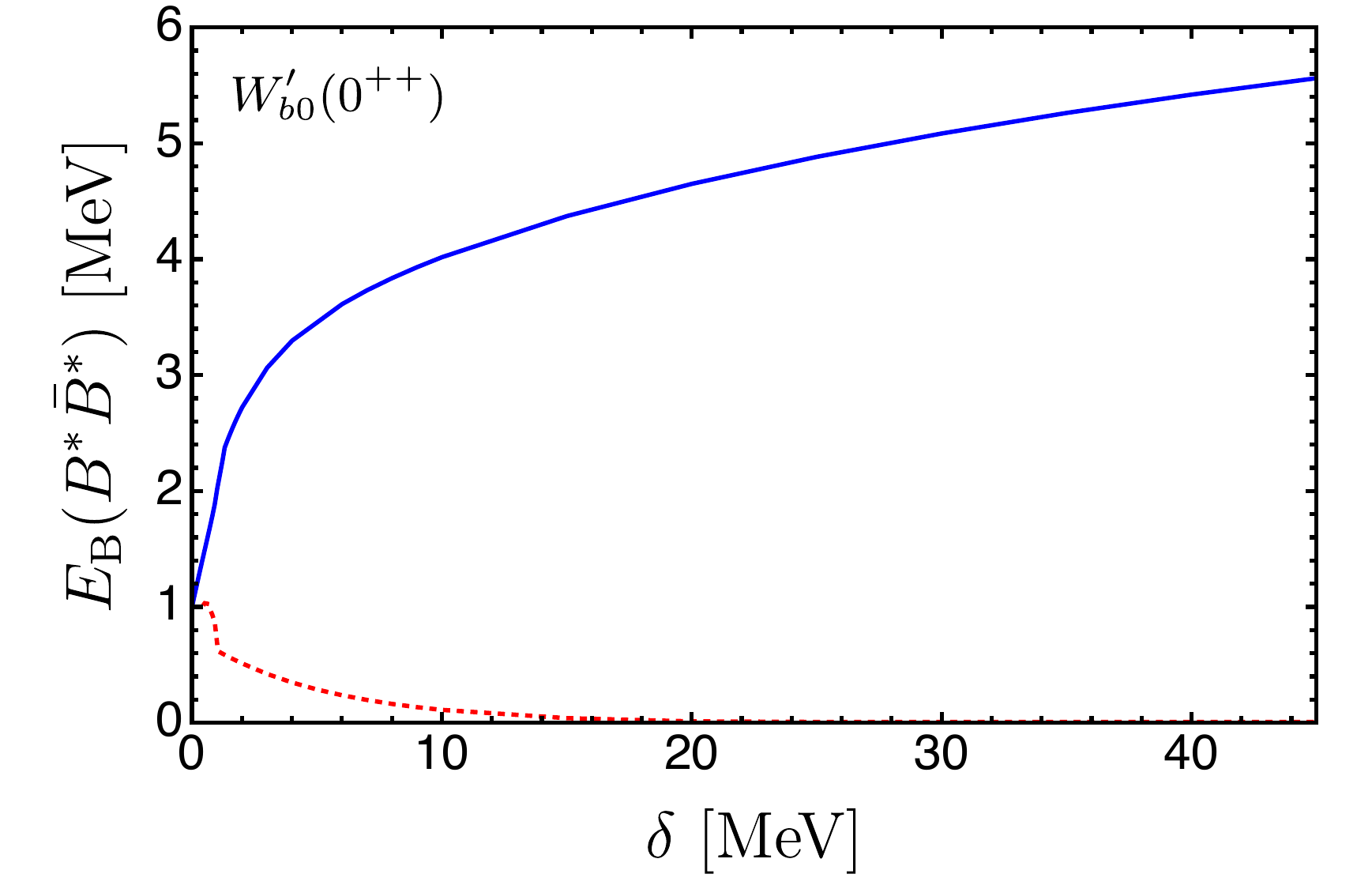}
\caption{
Binding energies of the $W_{b2} (2^{++})$ and $W'_{b0} (0^{++})$ states versus the mass splitting $\delta$ for two different solutions for the contact terms
in the pionless theory: the red dotted lines correspond to the first solution discussed in Sec.~\ref{sec:results} while the blue solid lines represent
the results of the alternative scenario with the interchanged contact terms (see the text for the further details). }
\label{fig:2branches}
\end{figure}

The two solutions presented in Eqs.~(\ref{HQSSlimit}) and (\ref{Sol.2}) and their evolution for finite $\delta$'s are different in a nontrivial way.
This is demonstrated in Fig.~\ref{fig:2branches} where, for definiteness, we used Eq.~(\ref{input1}) as input for the $Z_b$'s.
\footnote{Cusp-like structures in the binding energies seen in Figs.~\ref{fig:EBvsDelta} and \ref{fig:2branches} at small $\delta$'s appear when,
depending on the channel, the binding energy of the $Z_b'$ used as input coincides with $\delta$ 
(that is with the mass difference between the $B\bar{B}$ and $B\bar{B}^*$ or $B\bar{B}^*$ and $B^*\bar{B}^*$ thresholds)
or with $2\delta$ (the splitting between the $B\bar{B}$ and $B^*\bar{B}^*$ thresholds).}
However, as soon as the pion exchange is add\-ed, the strength of the off-diagonal elements
changes and the system gets sensitive to the interchange $C_1\leftrightarrow C_1'$.
It turns out that, in the presence of the OPE with its physical strength {(\ref{gc}), the $Z_b$'s binding energies can be fixed
to their physical values only for the first solution which in the strict HQSS limit corresponds to
Eq.~(\ref{HQSSlimit}) and which is discussed in the previous section}. Thus, the OPE naturally removes an ambiguity from the predictions for the spin partner 
states.

\section{Conclusions}
\label{sec:concl}

In this paper, we study the spin partners of the isovector bottomonium-like states $Z_b(10610)$ and $Z_b(10650)$ in the molecular model.
For definiteness, we assume both $Z_b$ states to be shallow bound states and treat their binding energies as input parameters
of the model. This allows
us to fix two low-energy constants describing the leading-order short-range interactions in the $1^{+-}$ channel.
Then HQSS is used to construct the contact
potentials for the
spin partner states of the $Z_b$ and $Z_b'$ with the quantum numbers $J^{++}$ ($J=0,1,2$), conventionally denoted as
$W_{bJ}$ ($J=0,1,2$). Motivated by the important role played by the nonperturbative pion dynamics in nuclear chiral
EFTs --- especially in the deuteron channel
which demonstrates certain similarities to the $Z_b$ states (see, for example, Refs.~\cite{Epelbaum:2008ga,Fleming:1999ee}) --- and
in the $c$-quark sector~\cite{Baru:2016iwj}, we performed
a fully nonperturbative calculation {with the one-pion exchange  
interaction added to the short-range terms. {In addition, we include one-$\eta$ exchange, also  
iterated to all orders for the sake of completeness,  to see if it indeed diminishes the effect of the OPE as claimed in the literature~\cite{Dias:2014pva}.}

 {Our results revealing  the influence of the OPE and OEE interactions on the properties of the bot\-to\-mo\-ni\-um-like systems under study
  can be summarised  as follows.}
\begin{itemize}
\item In contrast to the charmonium-like
systems --- see, for example, Refs.~\cite{Baru:2011rs,Baru:2016iwj} --- three-body effects in the bot\-to\-mo\-ni\-um sector play basically no role and,
in analogy with the $NN$ problem, they can be treated perturbatively. The reason why the static approximation for the OPE works well in the
bottomonium-like systems is twofold: first, the $B^*$-$B$ mass splitting is significantly smaller than the pion mass;
second, the masses of the $B$ mesons are large enough to make the typical three-body momenta
(proportional to $ \sqrt{m_{\pi}/m}$ $ \times $ soft scale --- see, for example, Refs.~\cite{Baru:2004kw,Baru:2009tx} for the details) sufficiently small compared to the
other relevant scales of the problem such as the binding momenta and the pion mass.
\item $D$ waves (particularly the $S$-$D$ transitions) play an important role and cannot be disregarded.
On the contrary, if only the central
$S$-wave part of the OPE interaction is retained, the problem is practically indistinguishable from the purely contact one \emph{after the parameters of the system are
re-adjusted} to place the $Z_b$ poles{, used as input,} to the prescribed locations. Since the leading non-trivial effect from the pions stems from 
the $S$-$D$ tensor forces,
the perturbative inclusion of the OPE 
is not sufficient.
\item In agreement with the results of Ref.~\cite{Baru:2016iwj}, the inclusion of the OPE interaction leads to selfconsistent results only
if all relevant partial waves as well as particle channels which are coupled via the pion-exchange potential are taken into account. Due to the larger
$b$-quark mass compared to the $c$-quark mass, consequences of uncontrolled omissions of partial waves are not that severe in the bottomonium
sector. However, {any partial neglect of this kind results in a strong cutoff dependence which reduces noticeably the predictive power of 
the approach.} 
\item The most important effect of the coupled-channel dynamics reveals itself in the
$2^{++}$ channel where, due to the $B^*\bar{B}^*\leftrightarrow B^{(*)}\bar{B}$ transitions driven by the
$S$-$D$ tensor forces from the OPE,
the corresponding partner state $W_{b2}$ acquires a width at the level of a few MeV
and it remains bound even in the limit of vanishing binding energies of the $Z_b$ states.

\item   {One-$\eta$ exchange,  iterated to all orders, plays a minor role for the problem at hand slightly diminishing the effect of the OPE. }
\end{itemize}

We, therefore, confirm the results of our previous studies of the OPE interaction which demonstrate its sizeable effect on the properties of the
near-threshold resonances --- see, for example, Refs.~\cite{Baru:2011rs, Baru:2016iwj}. On the other hand, our findings
suggest that the claim made in
Ref.~\cite{Voloshin:2015ypa} that the influence of the static $S$-wave OPE on the line shapes of the $Z_b$ states
can be as large as 30\% is not correct.
On the contrary, one expects that also for the line shapes the effect of the central $S$-wave OPE interaction
can be largely absorbed into re-definition of the parameters of the model. However, we delegate a detailed investigation of the
line shapes, which calls for an inclusion of the inelastic channels, to a later work.

 {The {predicted} masses of the spin partners of the $Z_b$ states are
quoted in Table~\ref{tab:res}. {In the same table we give the width of the $W_{b2}$ state due to the coupled-channel 
transitions to the open-flavour channels $B\bar B$ and $B\bar B^*$.}
From this table one can conclude that three sibling states with the quantum numbers $0^{++}$,
$1^{++}$, and $2^{++}$ are expected to exist near the $B\bar{B}, B\bar{B}^{*}$, and $B^*\bar{B}^*$ thresholds, respectively,
and to produce visible peak-like structures in the line shapes. A similar
$0^{++}$ state near the $B^*\bar{B}^*$ threshold has the pole located on a remote Riemann sheet.
It remains to be seen whether or not {this state} can be observed experimentally.}

{The probably most important finding of this work is that we predict the existence of an
isovector $2^{++}$ tensor state lying a few MeV below the $B^*\bar{B}^*$ threshold. Unlike its scalar partner states,  this tensor state should reveal itself
 as {\it a bound-state} peak in the $B^*\bar{B}^*$ line shape even in the limit of vanishing binding energies of the $Z_b$ states. }
 The width of this structure is expected at the level of a few MeV so in principle one should be able to resolve it
in the experiment. 

It should be stressed, however, that in this work we did not include any inelastic channels. The {$W_{b2}$} width
just reported --- see Table~\ref{tab:res} --- stems from the possible decays of the $W_{b2}$ to the $B\bar B^{(*)}$ final states in $D$ waves. 
Therefore, the {total} width
of this state should be somewhat bigger due to {its} decays to, for example, $\Upsilon(nS)\rho$, as proposed in Ref.~\cite{Bondar:2016hva}, or to
 {$\chi_{b1}\pi$ and $\chi_{b2}\pi$}. However, since inelastic channels play only a minor role for the $Z_b$'s, it is also natural to expect 
their contribution to be small enough for the $W_{b2}$ {which must make this state resolvable from the threshold and, therefore, 
detectable in the line shape}.

\acknowledgments

The authors are grateful to R. V. Mizuk for enlightening discussions and to F.-K. Guo for reading the manuscript and for valuable comments.
This work is supported in part by the DFG and the NSFC through funds provided to the Sino-German CRC 110 ``Symmetries
and the Emergence of Structure
in QCD'' (NSFC Grant No. 11261130311). Work of A. N. was performed within the Institute of Nuclear Physics and Engineering supported by
MEPhI Academic Excellence Project (contract No 02.a03.21.0005, 27.08.2013). He also acknowledges support from the
Russian Foundation for Basic Research (Grant No. 17-02-00485).
Work of V. B. is supported by the DFG (Grant No. GZ: BA 5443/1-1).

\providecommand{\href}[2]{#2}\begingroup\raggedright\endgroup

\end{document}